\documentclass[%
  pra,
superscriptaddress,
preprint,
 amsmath,amssymb,
 aps,
showkeys 
]{revtex4-1}
\usepackage[utf8]{inputenc}
\usepackage[english]{babel}
\usepackage{graphicx}% Include figure files
\usepackage{dcolumn}% Align table columns on decimal point
\usepackage{bm}% bold math
\usepackage{color}
\usepackage{hyperref}% add hypertext capabilities
\usepackage[normalem]{ulem} % add by ALine

\usepackage[mathlines]{lineno}% Enable numbering of text and display math
\makeatletter
\renewcommand{\fnum@figure}{\figurename {\bf \thefigure}}
\makeatother

\begin{document}
\renewcommand{\figurename}{{\bf Figure }}

\title{Shannon entropy of brain functional complex networks
\\  under the influence of the psychedelic Ayahuasca}

\author{A. Viol}
 \email{aline.viol@ufv.br}
 \affiliation {{Department of Physics, Universidade Federal do Rio Grande do Norte,} 59078-970 Natal--RN, Brazil}
\affiliation{Computational Biology Center,  T. J. Watson Research Center, IBM, 10598 Yorktown Heights--NY, USA}
\affiliation{{Department of Physics,  Universidade Federal de Viçosa,} 36570-000 Vi\c cosa--MG,Brazil} 
\author{Fernanda Palhano-Fontes}
\affiliation {Brain Institute, Universidade Federal do Rio Grande do Norte, 59078-970 Natal--RN, Brazil}

\author{Heloisa Onias}

\affiliation {Brain Institute, Universidade Federal do Rio Grande do Norte, 59078-970 Natal--RN, Brazil}

\author{Draulio B. de Araujo}%

\affiliation {Brain Institute, Universidade Federal do Rio Grande do Norte, 59078-970 Natal--RN, Brazil}
\author{G. M. Viswanathan}%
%\email{gandhi@fisica.ufrn.br}
\affiliation {{Department of Physics, Universidade Federal do Rio Grande do Norte,} 59078-970 Natal--RN, Brazil}

\affiliation {  National Institute of Science and Technology of Complex Systems 
Universidade Federal do Rio Grande do Norte, 59078-970 Natal--RN, Brazil}

\date{\today}% It is always \today, today,

\begin{abstract}
\hyphenchar\font=-1
The
entropic brain hypothesis holds that the key facts concerning 
psychedelics are  partially explained in
terms of increased entropy of the brain's functional connectivity.
  Ayahuasca is a psychedelic beverage of Amazonian indigenous origin with
legal status in Brazil
in religious and scientific settings.
In this context,
we use tools and concepts from the theory of complex networks to
analyze resting state fMRI data of the brains of human subjects
under two distinct conditions: (i) under ordinary waking state and
(ii) in an altered state of consciousness induced by ingestion of
Ayahuasca.
We report an increase in the Shannon entropy of the degree
distribution of the networks subsequent to Ayahuasca
ingestion.
We also find increased local and
decreased global network integration.
Our results are broadly consistent with the entropic brain
hypothesis.
Finally, we discuss our  findings in the context of descriptions of
``mind-expansion'' frequently seen in
self-reports of users of psychedelic drugs.

\end{abstract}

%\keywords{ Neuroscience, Complex Networks, fMRI, DMT, Shannon Entropy}
\maketitle

\hyphenchar\font=-1

Relatively little is known about how exactly psychedelics act on human
functional brain networks.  During the last few years, new
neuroimaging techniques, such as functional magnetic resonance imaging
(fMRI) \cite{buxton2009introduction,Heeger2002}, 
%,Logothetis2002}
%Raichle2009,Buxton2013, Buxton1997} 
have allowed noninvasive investigation of global brain activity in a
variety of conditions, e.g., under anaesthesia, sleep, coma, and in
altered states of consciousness induced by psychedelic drugs
\cite{Schrouff2011,schoter2012, Noirhomme2010,
  Andrade2011,draulio2012,
  carhatharris2012,Carhart-Harris2016,fernanda2014}.
%Edlow2013, lizette2012, Fransson2009}.
  %meditation \cite{brewer2011}
%
Recently, Carhart-Harris {\it et al.}  proposed a hypothesis known as
the {\it entropic brain}, which holds that the stylized facts
concerning altered states of consciousness induced by psychedelics can
be partially explained in terms of higher entropy of the brain’s
functional connectivity \cite{ Carhart-Harris2014}. Although the
entropy of the brain has never been directly measured, the entropic
brain hypothesis is empirically supported by several recent
studies.
For example, Sarasso {\it et al.} have reported complex spatiotemporal
cortical activation pattern during anesthesia with ketamine, which can
induce vivid experiences (``ketamine dreams'') \cite{sarasso-xxx}.
Similarly,
Petri {\it et al.} found that after administration of the psychedelic
psilocybin, the brain's functional patterns undergo a dramatic
change characterized by the appearance of many transient low-stability
structures~\cite{petri-xxx}.
Perhaps the most convincing evidence supporting the hypothesis
thus far 
has
come from  the study  undertaken by Tagliazucchi {\it et al.}
\cite{tagliazicchi2014}, who reported a larger repertoire of brain
dynamical states during the psychedelic experience with psilocybin.
They inferred an increase in the entropy of the functional
connectivity in several regions of the brain, by studying the temporal
evolution (i.e., dynamics) of the connectivity graphs.
Here we directly measure increases in entropy associated with the
functional connectivity of the whole brain under the influence of a
psychedelic.  Specifically, we analyze fMRI functional connectivity of
human subjects before and after they ingest the psychoactive brew
Ayahuasca and report an increase in the Shannon entropy.
  This is the first time that the entropy of the functional networks
  of the human brain has been directly measured in altered states of
  mind on a global scale, i.e. considering the entire brain.

\hyphenchar\font=`\-
  
Ayahuasca is a beverage of Amazonian indigenous origin and has legal
status in Brazil in religious and scientific settings
\cite{labate2014prohibition}.
It contains the powerful psychedelic $N,N$- dimethyltryptamine (DMT),
together with harmala alkaloids that are known to be monoamine oxidase
inhibitors (MAOIs).
The beverage is typically obtained by decoction of two plants from the Amazonian
flora: the bush {\it Psychotria viridis}, that contains DMT, and the
liana {\it Banisteriopsis caapi}, that contains MAOIs
\cite{mckenna2004}.  DMT is usually rapidly metabolized by monoamine
oxidase (MAO), but the presence of MAOI allows DMT to cross the
blood-brain barrier and to exert its effects
%>>>>>GANDHI, AQUI SÃO EXATAMENTE AS MESMAS CITAÇÕES NA MESMA FRASE. ISSO ESTÁ OK?<<<<<<<<<<<<<<<<<<<<<<<<<<<<<<
\cite{shanon2002antipodes,huxley2004doors,hofmann1983lsd,Hollister1962235,
  grof1980lsd,Griffiths2006}. Similar to LSD, mescaline and psilocybin
\cite{shanon2002antipodes,huxley2004doors,hofmann1983lsd,Hollister1962235,
  grof1980lsd,Griffiths2006}, Ayahuasca can cause profound changes of
perception and cognition, with
users reporting increase of awareness, flexible thoughts, insights,
disintegration of the self, and attentiveness
\cite{shanon2002antipodes,riba2001}.
There is growing interest in Ayahuasca, partially due to recent
findings showing that it may be effective in treating mental
disorders, such as depression \cite{Sanches2016} and behavioral
addiction \cite{OSORIO2015,labate2013therapeutic,Nunes2016}.
Similar therapeutic potential has also been pointed out for other psychedelics
\cite{krebs2012lysergic,grof1980lsd,johnson2014pilot,albaugh1974peyote,
frederking1955intoxicant, Carhart-Harris2016}. 

\hyphenchar\font=-1

For analysis we use tools and concepts from the field of
complex networks, a brief history of which follows.  The application
of graph theory to phase transitions and complex systems led to
significant progress in understanding a variety of cooperative
phenomena over a period of several decades.
In the 1960s, the books by Harary,
%
%\cite{harary1969graph,harary1967graph}
%
especially {\it Graph Theory and Theoretical
  Physics}~\cite{harary1967graph}, introduced readers to powerful
mathematical techniques. The chapter by Kastelyn, still considered to
be a classic, showed that difficult combinatorial problems of exact
enumeration could be attacked via graph theory, including the exact
solution of the two-dimensional Ising model (e.g., see
Feynman~\cite{feynman1998statistical}).
In the 1980s, certain families of neural network  
models were shown to be equivalent to Ising systems, e.g., the
Hopfield network \cite{hopfield1982} is a content-addressable memory
which is isomorphic to a generalized Ising
model~\cite{amit1992modeling}.
Beginning in the 1990s, new approaches to networks, giving emphasis to
concepts such as the node degree distribution, clustering,
assortativity, small-worldliness and network efficiencies, led
eventually to what has become {the new field of complex networks
\cite{Barabasi2002,newman2010networks}. 
%  WS1998,
%latora2001
%  newman2002}.
%
These new tools and concepts}
\cite{ Albert02, 
Bornholdt2003, caldarelli2013scale}
%Boccaletti2006175} 
%pastor2007evolution,
 % barrat2008dynamical,
 % Bullmore2009, brain net
 % schoter2012, 
  %latora2003}
%
have found successful application in the study of diverse phenomena,
such as air transportation networks
\cite{Verma2014}, 
% Grady2012,Guimera2005,},
%
terrorist networks \cite{terror}, 
gene regulatory networks \cite{Magtanong2011a},
and functional brain networks
\cite{sporns2016networks,Bullmore2009,
mckenna1994,rubinov2010,meunier2010}.
%achard2006,
%Bassett2010a,
%,schoter2012,liuyong2008}.
%
% Indeed,
% the human brain is arguably one of the most complex networks
%
We approach the human brain from this perspective of {complex networks}
\cite{braincomplexsystem2,braincomplexsystem3}. %braincomplexsystem1}.
% Here we review only what is necessary for our purposes. 
%
%
%The fundamental concepts of complex networks comes from graph theory.

%Our dataset consists of fMRI images of the entire brain.
Ten healthy volunteers were submitted to two distinct scanning
sessions: (i) before and (ii) 40 minutes after Ayahuasca intake, when
the subjective effects are noticeable.  In both cases, participants
were instructed to close their eyes and remain awake and at rest,
without performing any task.
%
% We thus have one dataset acquired from subjects under the influence of
% Ayahuasca, as well as a second control dataset for the same subjects
% in the ordinary state of wakeful rest.
%
We performed a standard preprocessing on all samples of the fMRI
data (see Methods for details concerning data acquisition and
preprocessing).

\pagebreak

\begin{figure}[p]
   \noindent (a)\hspace{2.7cm} Before \hspace{7.35cm} After
   \hfill ~
   \vspace{-4mm}
   \begin{center}
\parbox{15cm}{
     \hspace{-0.0cm}\includegraphics[height=4.1cm]{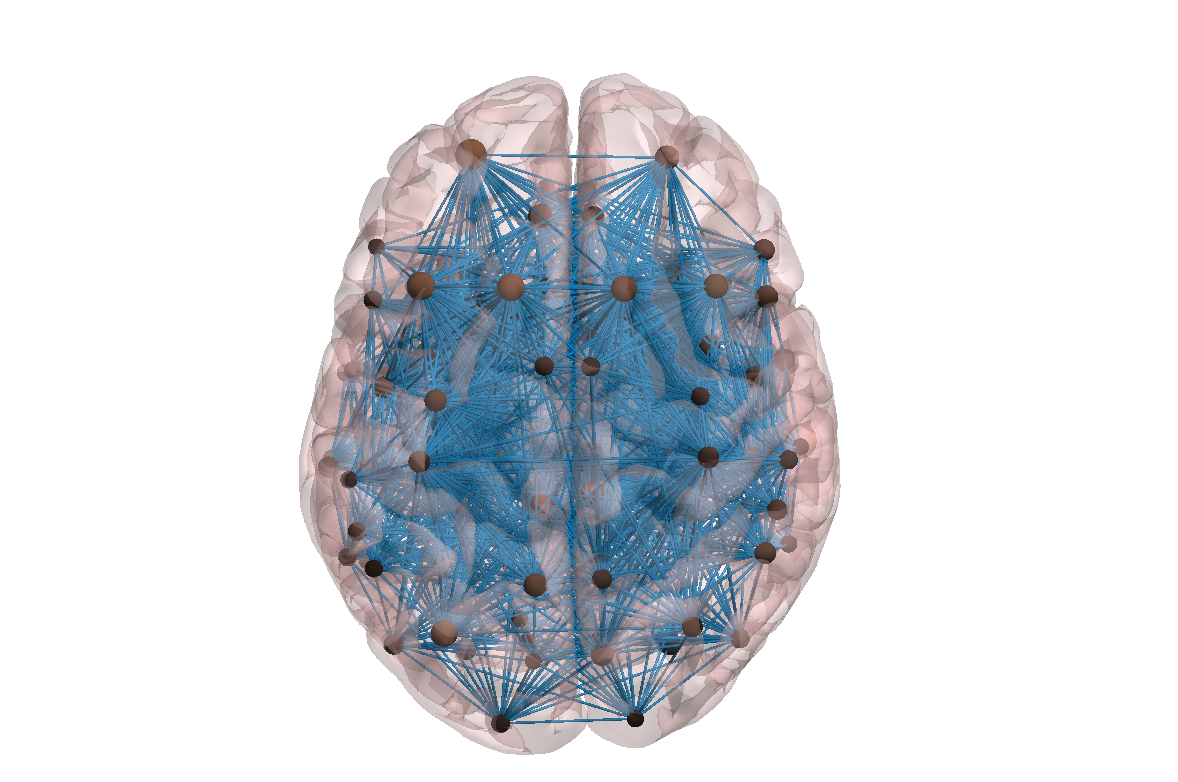}
    \hspace{2.2cm}\includegraphics[height=4.1cm]{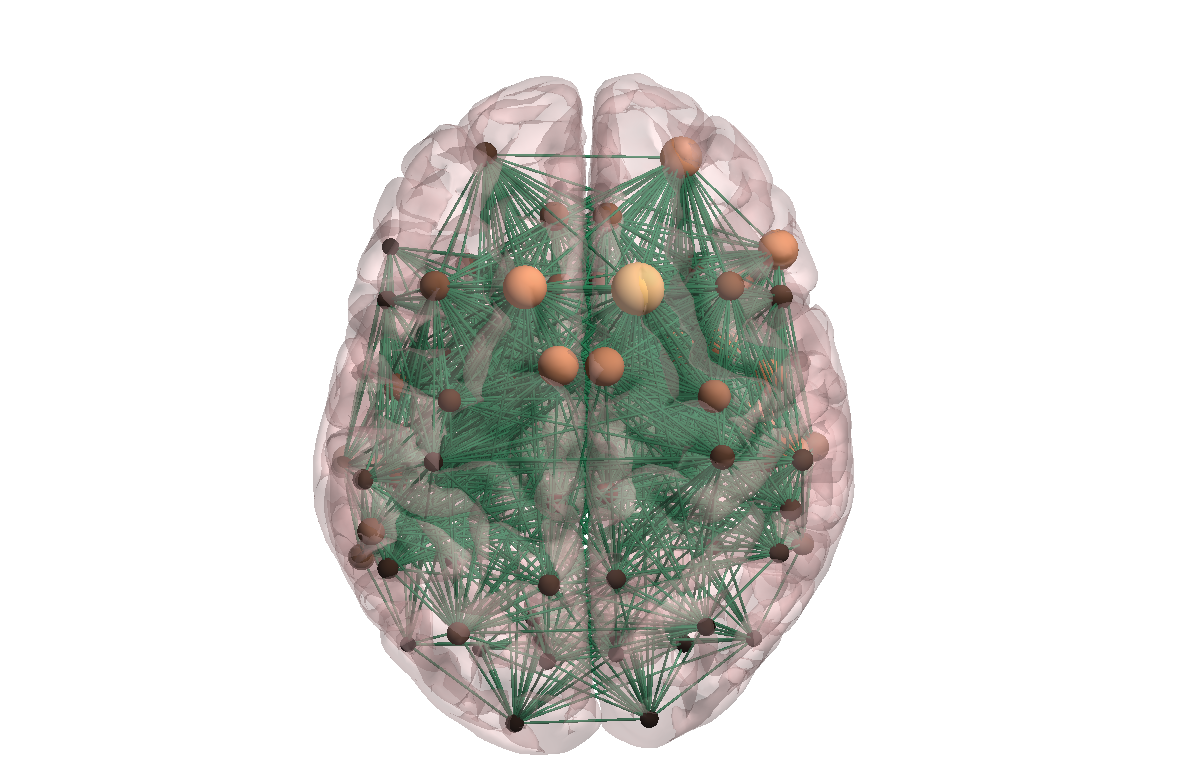} \\
%  \hspace{-1.0cm}\includegraphics[height=0.2]{figuras/sagital_s8_before.png}
%  \hspace{-1.0cm}\includegraphics[height=0.2]{figuras/sagital_s8_after.png} \\
  \hspace{-0cm}\includegraphics[height=3.4cm]{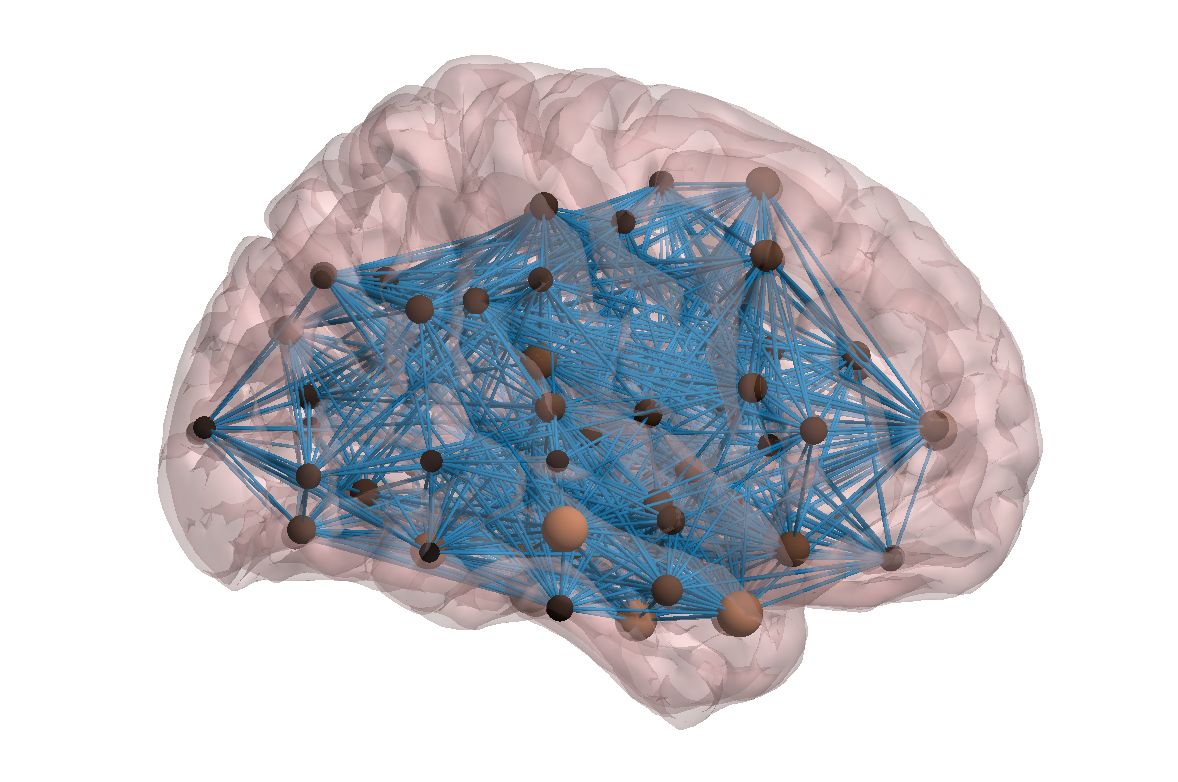}
 \hspace{3.4cm}\includegraphics[height=3.4cm]{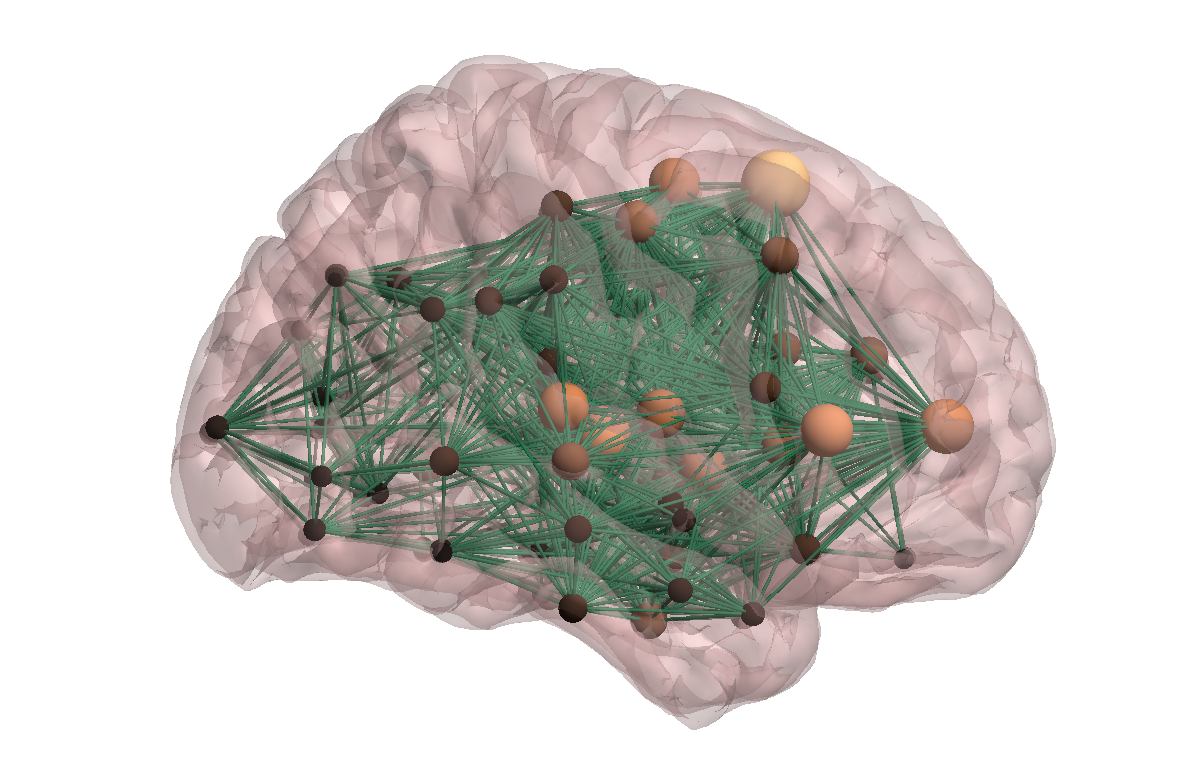} \\
   \hspace{-00cm}\includegraphics[height=3.5cm]{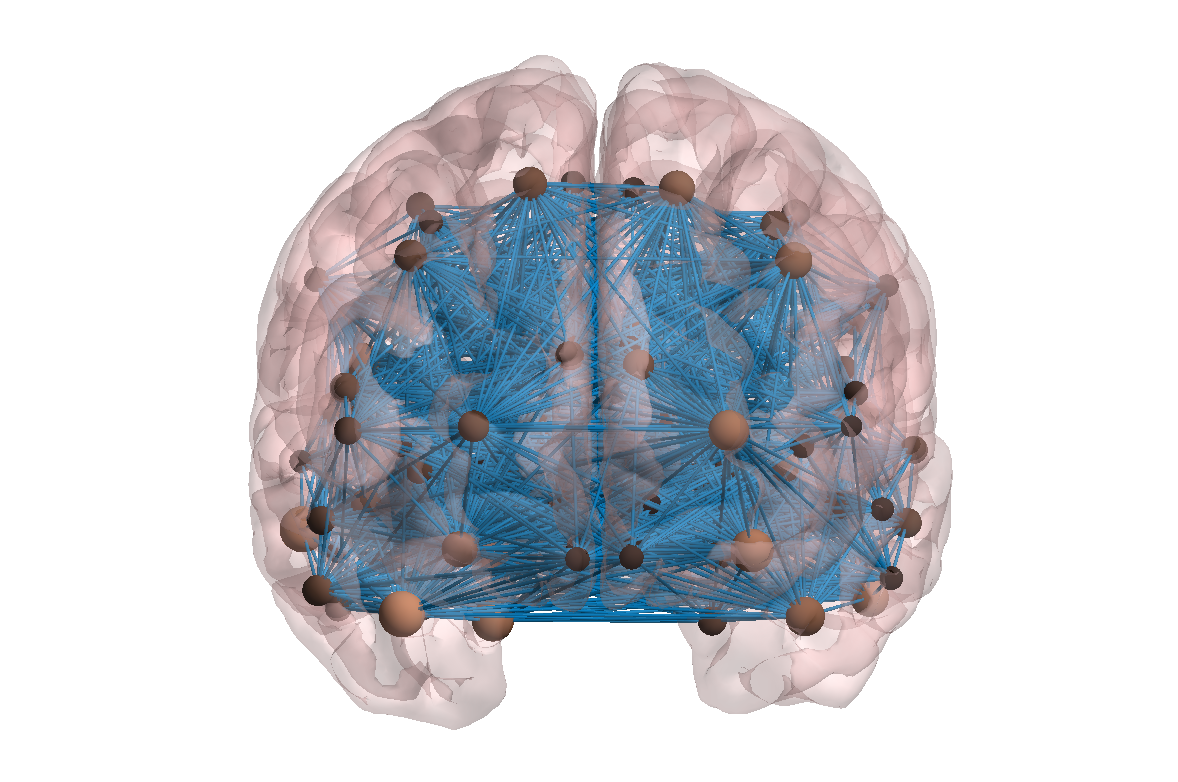}
   \hspace{3.4cm}\includegraphics[height=3.5cm]{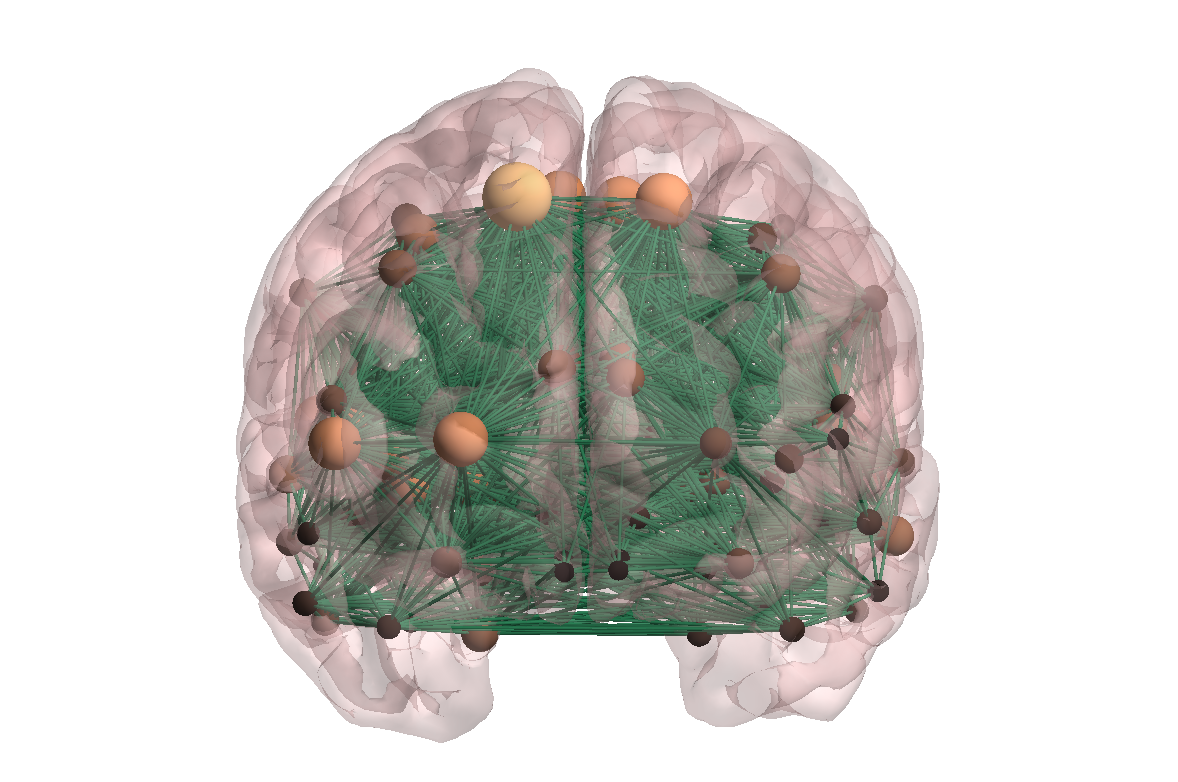}\\
%
%   \hspace{1cm}(a) \hspace{2in}(b)\\
}
\end{center}
   \noindent (b) \hfill ~

\vspace{-1mm}
\centerline{     \includegraphics[scale=0.43]{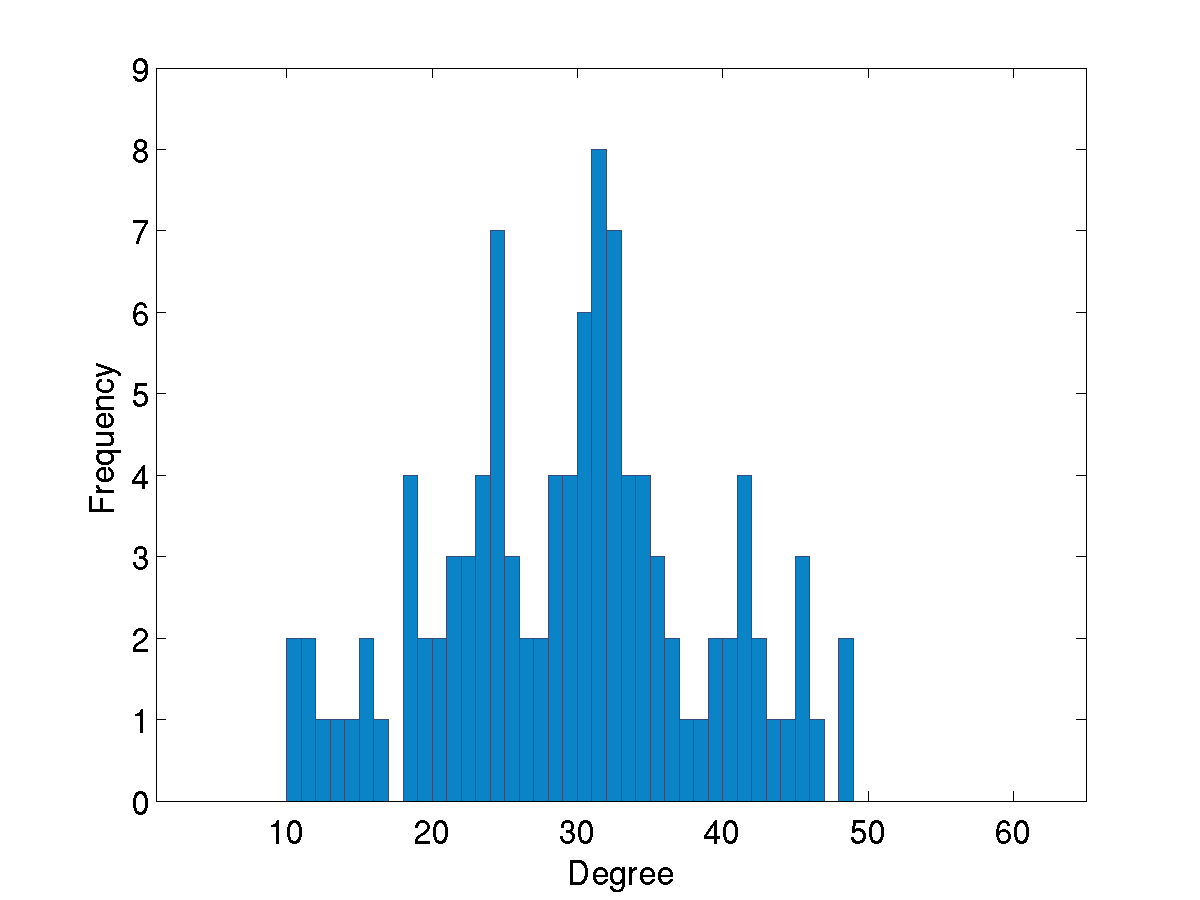}   
\!\!\!\!\!\!\!\!\!\!\!\!\!\!\!
\includegraphics[scale=0.43]{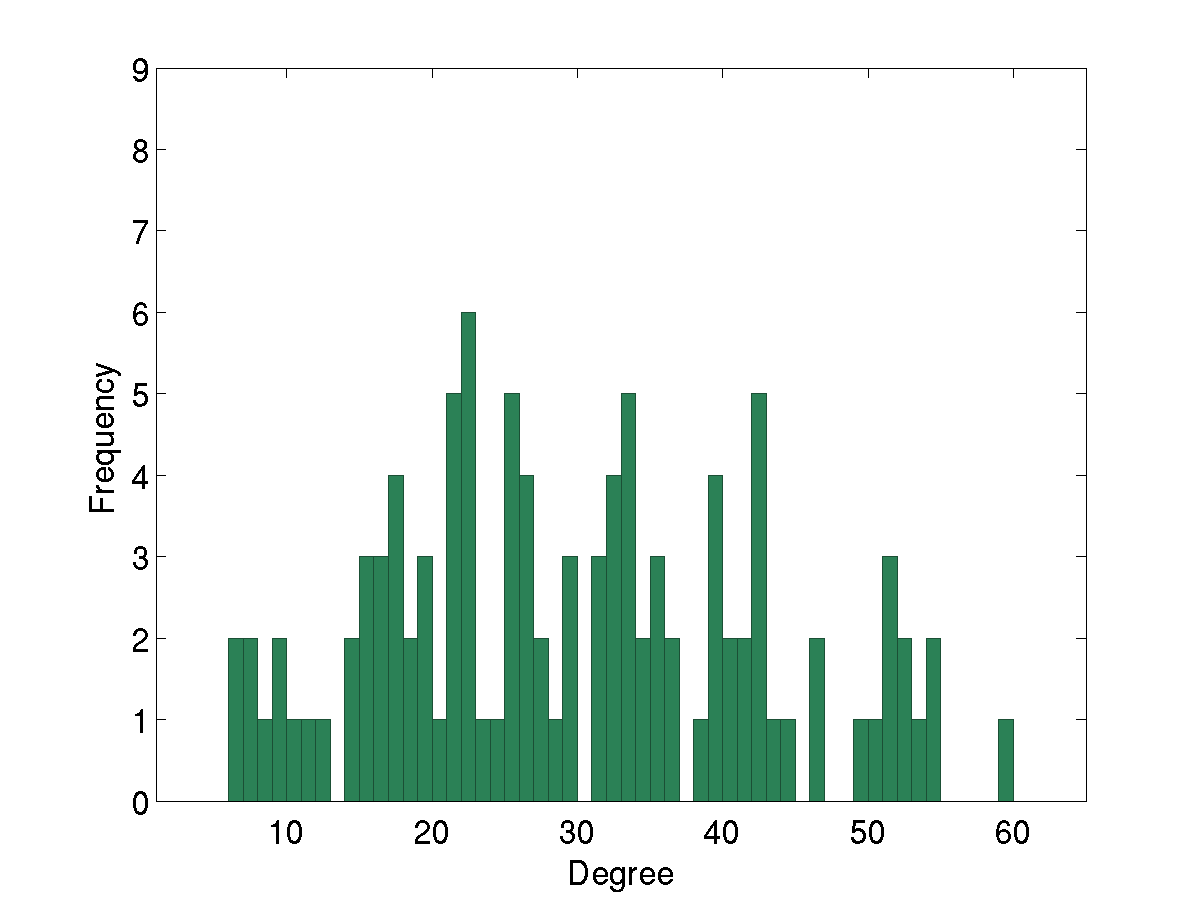}}

 %
% \end{center}
\caption{{\bf Illustrative example of functional brain networks.} (a) 3 views of a complex network generated from brain fMRI
  data of one of the subjects, before (left) and after (right)
  Ayahuasca ingestion (mean node degree $ \langle k \rangle=30$).  The
  spheres represent nodes and sphere size is proportional to the node
  degree.  (b) histograms of the node degrees, corresponding to the
  networks shown in (a).  After Ayahuasca intake, the distribution is
  wider, indicating a higher entropy.
In (a) we have used the BrainNet Viewer
(http://www.nitrc.org/projects/bnv) for visualization.  }
  \label{braindegree}
  \end{figure}

Data analysis consists of two main steps.  In the first step,
we use fMRI data to generate complex networks to  represent the actual
functional brain connectivity patterns.
In the second step, we use the networks generated in step
1 as inputs and calculate network characteristics as output,
using techniques from the theory of complex networks.
%
% including standard quantities such as the clustering coefficient,
% geodesic distance, local and global efficiencies.  
The Methods section describes both steps in detail.
%We also
%give a brief review of network concepts.
%
More information 
about most of the methods used here can be found in
refs.~\cite{onias2014,schoter2012, liuyong2008}.
  Figure \ref{braindegree} shows the networks generated from one
  subject before and after Ayahuasca intake, for one specific choice
  of mean node degree.  The spheres represent nodes, with sphere size
  proportional to the degree of the node.  The lower plots show
  histograms of node degrees.

The main result that we report here is an increase in the Shannon
entropy of the degree distribution for the functional brain networks
subsequent to Ayahuasca ingestion.
We also find that the geodesic distance increases during the effects of Ayahuasca, i.e. qualitatively the network becomes
``larger.''
More generally, we also find that these functional brain networks become
less connected globally but more connected locally.

The key technical innovation is the measurement of the Shannon entropy
of the degree distribution of the complex networks that represent the
functional connectivity of the human brain.  This novel use of the
Shannon entropy allows the brain to be studied from the perspective of
information theory in a manner previously unexploited.
Moreover, the Shannon entropy is also very closely related to the
Boltzmann-Gibbs entropy used in statistical mechanics.  Hence, our
approach to studying the brain experimentally is grounded in two
strong theoretical traditions: graph theory and complex networks
on the one hand,  and information theory and statistical physics on
the other.  { Our study also represents a significant advance for the
  following additional reasons:
(i) our results unveil how Ayahuasca (and likely most other tryptamine psychedelics) alter brain function,
both locally and globally; 
%
%(ii)
%it is the first time that the Shannon entropy has been applied to measure disorder in complex networks generally,
% and functional  brain networks specifically; and 
%
(ii) 
%We are the fisrt to propose the shannon entropy of degree distribution to caracterize functional networks of altereted state of consciencioness. 
it is the first time that
this specific approach
has been applied to characterize functional brain
networks in altered states of consciousness; (iii) our study of
Ayahuasca covers all brain regions; and (vi) the method we have developed can be
immediately applied to study a variety of other phenomena (e.g., the
effects of medication for mental health disorders). }

  \bigskip
  \bigskip

\section*{Results} \label{sec-results}
%  \bigskip  \bigskip 
%  \noindent {\bf RESULTS} 

\subsection*{Increase of the Shannon entropy of the degree distributions}

\begin{figure}[h]
  \begin{center}
%    \hspace{0cm}\includegraphics[scale=0.6]{figuras/pdf.png}
  \hspace{.0cm}(a)\includegraphics[scale=0.41]{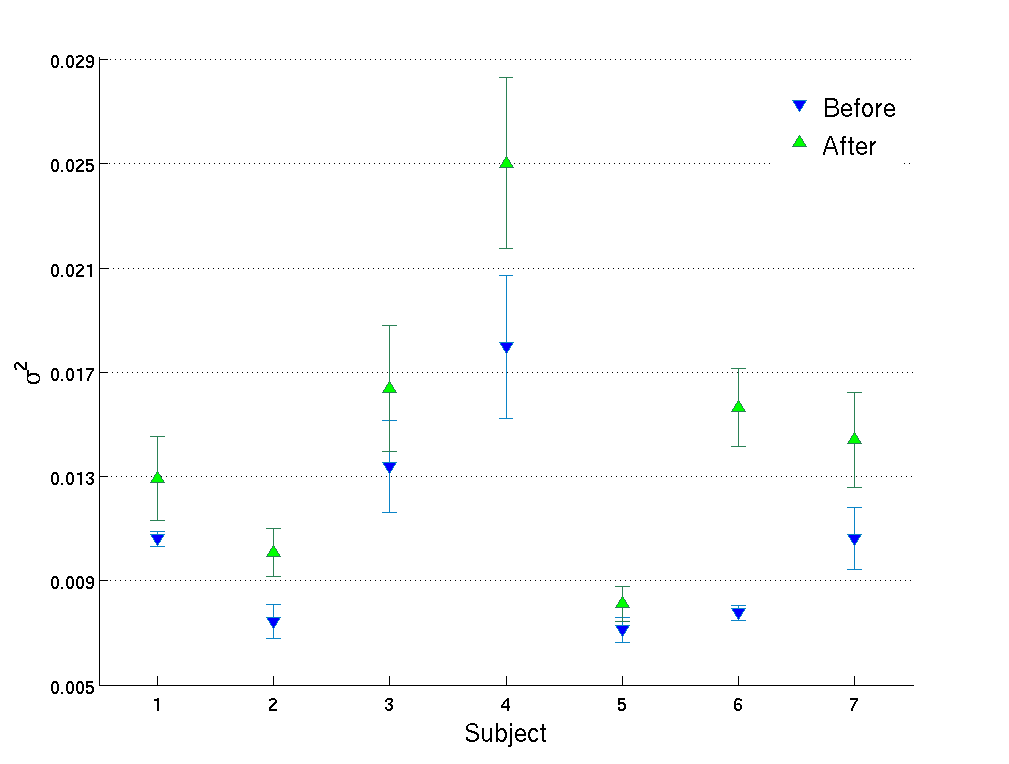}\\
    \hspace{.0cm}(b)\includegraphics[scale=0.41]{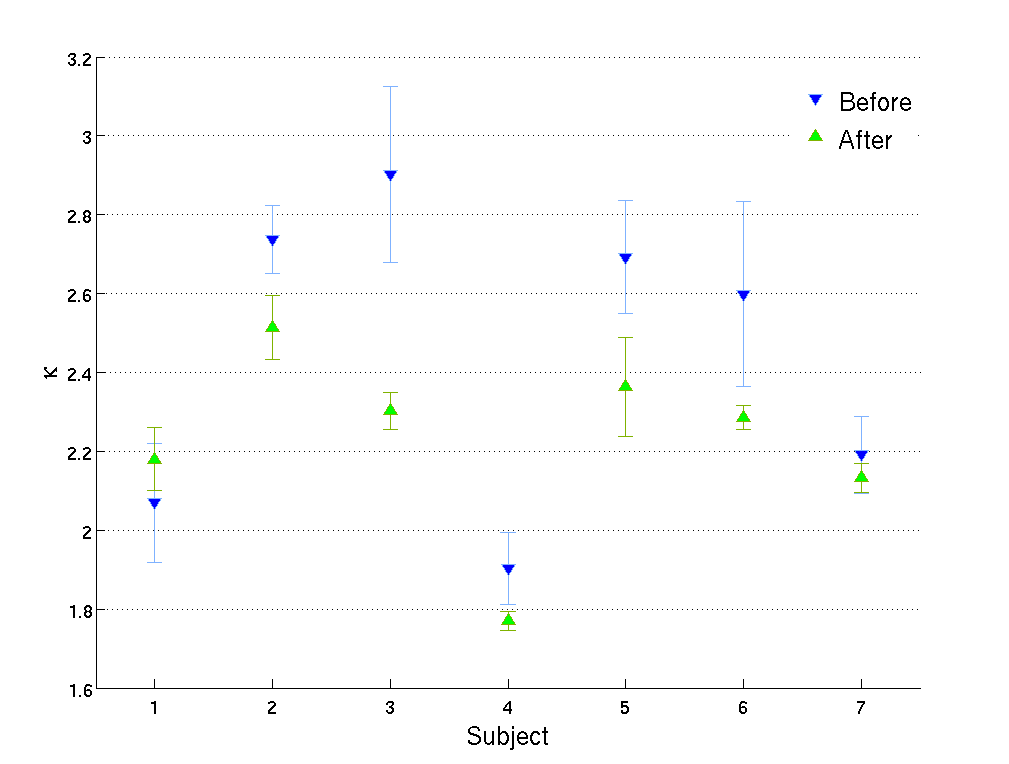}
  
    \caption{{\bf Variance and kurtosis of the degree distribution.} Mean $\pm$ 1 standard deviation calculated over all 16
      networks of the degree variance (a) and kurtosis (b),
      shown for each subject  (blue $\triangledown$) and after (green $\triangle$) Ayahuasca ingestion.  The individual values for the
      degree variance and kurtosis are calculated separately for each
      network.  We find higher variance and (mostly) lower kurtosis
      after Ayahuasca, hence the node distributions change shape and
      become less ``peaked.''  Such behavior is again consistent with (if
      not suggestive of) a higher Shannon entropy after Ayahuasca.}
  \label{moment}
  \end{center}
  \end{figure}

  We find evidence of  significant changes in  the functional brain networks of
  subjects before and after ingestion of Ayahuasca.
  Figure \ref{moment} shows 2nd as well as 4th central moments of the
  degree distributions for each subject. The individual values are
  calculated separately for each network. We find an increase of
  variance for all subjects after Ayahuasca intake and a decrease of
  kurtosis for almost all of them (6 subjects).  These findings
  indicate that the degree distributions become less peaked and wider.
  This behavior is suggestive of an increase of the Shannon entropy
  for the degree distributions after Ayahuasca ingestion.

\begin{figure}[h]
   \begin{center}

     ~
     \bigskip     \bigskip      \bigskip 
     
  \centerline{\includegraphics[scale=0.6]{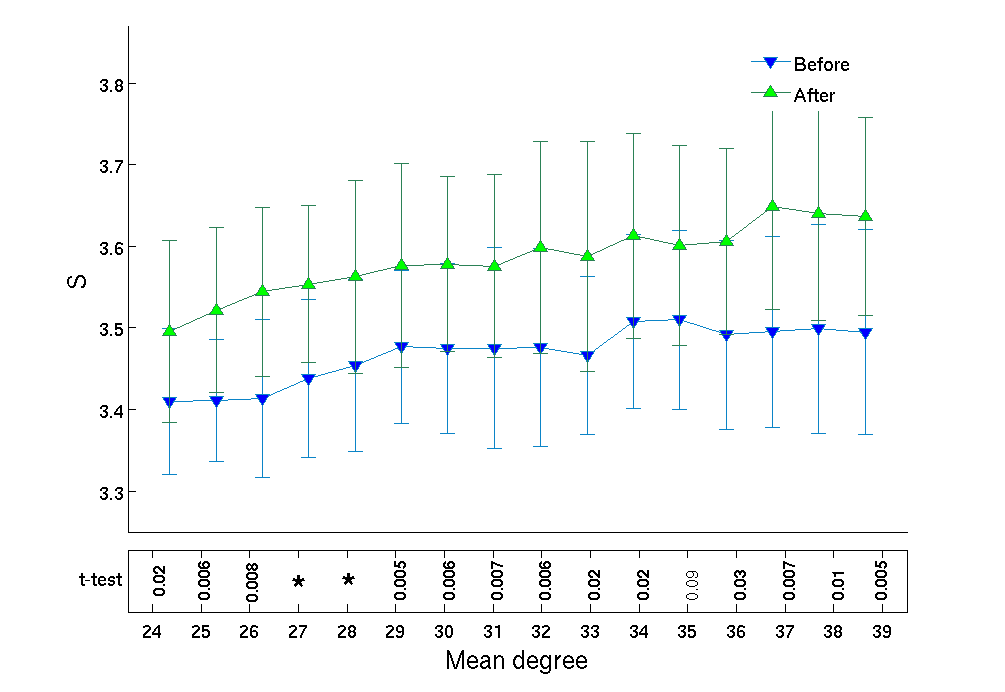} }

  \caption{{\bf Entropy grows after
      Ayahuasca ingestion.} Mean $\pm$ 1 standard deviation of the
    Shannon entropy of the distribution of node degrees, calculated
    over all 7 subjects, as a function of mean degree $k$, before
    (blue $\triangledown$) and after (green $\triangle$) Ayahuasca
    intake.  The bottom row lists $p$-values for Student's paired
    $t$-test, with values $p<0.005$ indicated by asterisks (*). We
    thus see evidence against the null hypothesis of no change in
    entropy.  Indeed, we find a significant increase in the entropy of
    the degree distributions after Ayahuasca ingestion. This entropy
    increase is the main result that we report.}
%[h p]=ttest(data(before,:)',data(after,:)')
    % correct.
%h = ttest(x,y) returns a test decision for the null hypothesis that the data in x – y comes from a normal distribution with mean equal to zero and unknown variance, using the paired-sample t-test.
      
  \label{entropy}
  \end{center}
  \end{figure}

\begin{figure}[h]
    (a) \hfill ~

    \vspace{-1cm}
    \centerline{\includegraphics[scale=0.5]{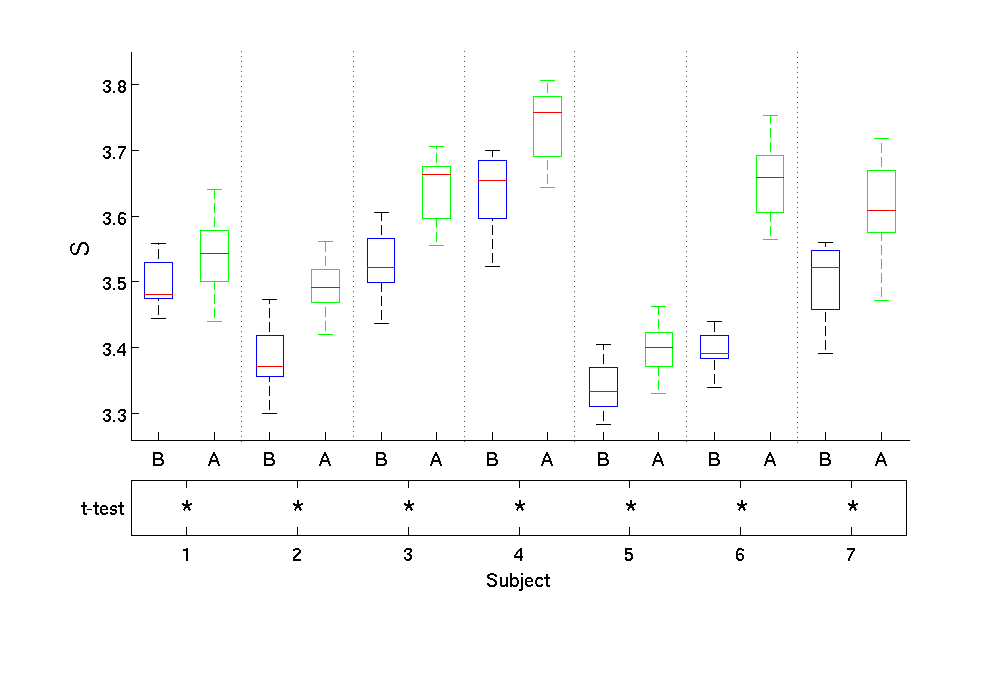}}

\vspace{-7mm}

    (b)\hfill ~
    \vspace{-1cm}

\centerline{\includegraphics[scale=0.5]{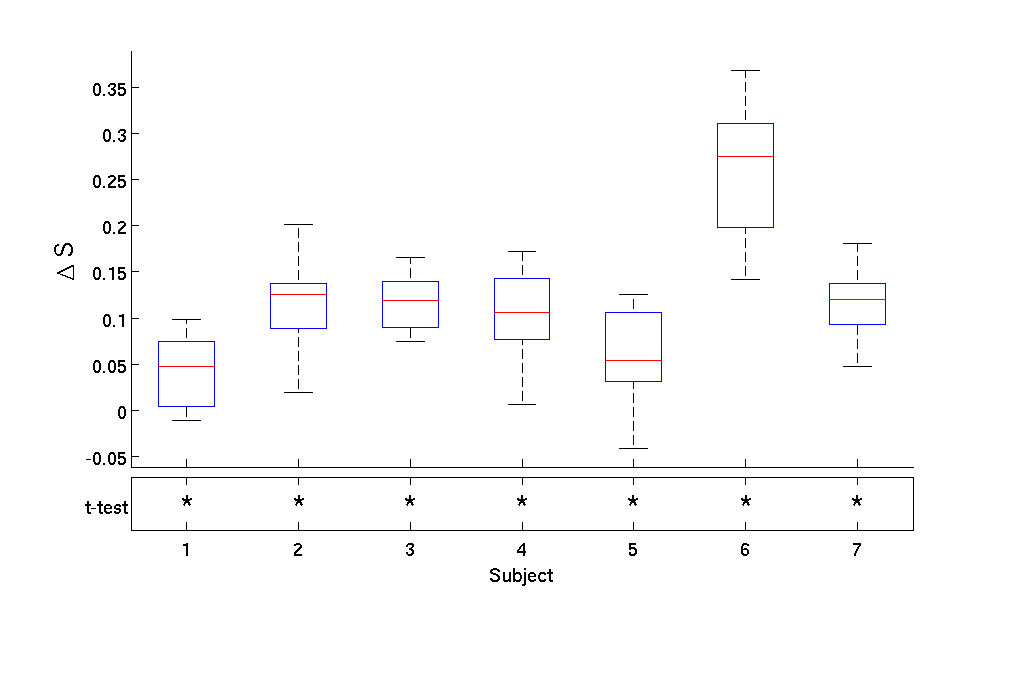}}

\vspace{-10mm}
\caption{{\bf Entropy growth per subject.} (a) Boxplot of the entropy distribution and before (B) and
    after (A) Ayahuasca ingestion and (b) boxplot of entropy increase,
    for all 7 subjects.  Note the significant increase in entropy
    after Ayahuasca ingestion. There are 16 values of entropy per
    subject, as discussed in the text.  The bars show minimum and
    maximum values and the box shows the 2nd and 3th quartiles, with
    the median shown dividing the box (in red).  The asterisks (*) in
    the bottom rows in both plots indicate $p$-values $p<0.005$ for
    Student's paired $t$-test in (a) and $t$-test for zero mean in
    (b). Subject-by-subject, we thus find  strong evidence against the
    null hypothesis of no entropy change.}
  \label{boxentropy}
  \end{figure}

 Figure \ref{entropy} shows the average Shannon entropy of the degree
 distributions as a function of mean degree, considering networks from
 all subjects, before and after Ayahuasca intake.
%
% The figure consider distinct sets of networks with identical mean degree, to
% allow fair comparisons between before and after Ayahuasca intake.
A fair comparison of the ``before'' and ``after'' networks is possible
by considering the entropy of networks of identical mean degree.
We find an increase in the entropy of the degree distributions
after Ayahuasca ingestion.  In order to better evaluate the
consistency of this result, we also calculate the average Shannon
entropy subject-by-subject, before and after Ayahuasca (Figure
\ref{boxentropy}). We find significant increased entropy for all
individual subjects.

\subsection*{Iso-entropic randomized networks}

The degree distribution does not completely define a network, however
it can have great influence over other network properties.
One can quantify this influence by comparing any given network $G$ to
other networks chosen randomly from the ensemble of networks that have
exactly the same degree distribution.
We refer to such networks  as ``randomized networks.''
By definition, all such randomized networks have the same entropy as
the original network $G$, i.e. they are 
iso-entropic to $G$.

An efficient way of generating such randomized networks is the
Maslov algorithm \cite{Maslov2002} (see Methods).
Whereas entropy is conserved by the Maslov algorithm, the clustering
coefficient, geodesic distances and efficiencies are not.  By
comparing these non-conserved quantities before and after
randomization, we can distinguish effects that are due solely to
changes in the degree distribution from those that are sensitive to
how links are more specifically arranged.

We generate a set of 30 iso-entropic randomized networks for each
original network, for all subjects both before and after Ayahuasca
ingestion.  Comparison of the original networks with the randomized
networks yields important information concerning to what degree the
changes in quantities such as geodesic distance, clustering
coefficients, and global and local efficiencies can be accounted for by
the changes in the degree distributions (see results described below).

\subsection*{Decrease of global integration}

 Figure \ref{distance} shows an increase of mean geodesic distance and a
decrease of global efficiency after Ayahuasca ingestion.
To determine how much of the change in geodesic distance is due to the
change in the degree distribution, we also calculated the geodesic
distance and global efficiency for the iso-entropic randomized
networks.
Note how the values for those networks are quite different
compared to the non-randomized networks.  We conclude that the change
in degree distribution cannot explain the entire change in geodesic
distance.  
The inset in the middle panels shows the change in the normalized mean
geodesic distance and global efficiency, which we define as the ratio
$D/D_{\mbox{\tiny rand}}$ and similarly for the global efficiency (see
\cite{Maslov2002,schoter2012}) . We see, indeed, that these ratios are
not close to zero. If the change in degree distribution could account
for all the change in geodesic distance and efficiency, then the
change in these ratios would be close to zero. Significant changes are
also observed at the individual level and are again consistent for
all subjects (Figure \ref{distance} (e) and \ref{distance} (f)).

\subsection*{Increase of local integration}

Figure \ref{clustering} shows an increase of clustering coefficients
and local efficiency after Ayahuasca ingestion. In contrast to the
behavior of the metrics discussed above, almost identical changes are
seen for iso-entropic networks. This result indicates that the
variation in degree distribution can account for most of the change in
clustering and local efficiency.  The insets in the middle panel show
the change in the normalized clustering and local efficiency, which we
define as the ratio $C/C_{\mbox{\tiny rand}}$ and similarly for the
local efficiency (see \cite{Maslov2002,schoter2012}). We see, indeed,
that these ratios are close to zero.

%{\color{red}

\section*{Discussion}

Our results reveal some remarkable findings, the most important of
which is that the entropy increases after Ayahuasca ingestion.  The
following also increase: geodesic distance, clustering coefficient and
local efficiency. However, the global efficiency decreases.  Overall,
we find an increase of local integration and a decrease of global
integration in the functional brain networks.

We interpret these findings in the context of some
well understood prototypical classes of networks.
Regular lattices have fixed coordination number, hence all
nodes have the same degree and the Shannon entropy of the degree
distribution is thus zero.
%
  %\color{blue}
%\sout{Entropy increases in the node degree distribution are thus seen
%when regular networks such as lattices are randomized.} 
In contrast, the entropy is high in networks with broad distributions of degree.
%\sout{ In contrast, the observed increase in clustering is more usually associated with
%decrease disorder.}
In the context of the Watts-Strogatz model \cite{WS1998}, clustering
and geodesic distance both decrease when highly regular networks are
transformed into small-world networks by randomly re-assigning the
links.
%
%\textcolor{red}
  {Whereas clustering and geodesic distances decrease
  with increasing randomness in such models, we find the opposite
  behavior for Ayahuasca, i.e., randomness as measured by the Shannon
  entropy of the node degree distribution increases in parallel with
  clustering and geodesic distances.}
Hence, our findings cannot be reduced to simple explanations of greater or
lesser randomness.
Locally, there is an increase in
integration (as measured by network efficiency), but globally there is
a decrease in integration.
Indeed the increase of geodesic distance and decrease of global
efficiency after Ayahuasca intake signify that the functional brain networks are less globally integrated.
One possible interpretation of these findings is that the increase of
local robustness and the decrease of global integration reflect a variation 
in modular structure of the network.
Recent studies have reported the presence of modularity in functional
brain networks on several scales \cite{ferrarini2009, meunier2010,
  Nicolini2016}.  Modular networks are characterized by the existence
of reasonably well-defined subnetworks in which internal connections
are denser than connections between  distinct subnetworks
\cite{meunier2010}.
However, traditional algorithms \cite{Newman2004, Guimera2005a,
 Blondel2008} were not able to detect variation on modular
structure features between our sets of networks.

Our results are broadly consistent with the entropic brain hypothesis,
hence we discuss the latter in the context of our findings.  The
hypothesis maintains that the mental state induced by psychedelics,
which the original authors 
term ``primary-state,'' presents relatively elevated entropy
in some features of brain organization, compared to the ordinary
waking state (termed ``secondary'') \cite{Carhart-Harris2014}.
Although it may be somewhat counter-intuitive that the psychedelic
state is considered primary while ordinary consciousness is secondary,
their hypothesis is inherently plausible considering that a wider
spectrum of experiences is possible with psychedelics than in ordinary
consciousness.  In this sense, ordinary consciousness can be thought
of as a restriction or constrained special case of a more primary
consciousness.  The  hypothesized lower entropy of ordinary
consciousness relative to primary consciousness is attributed to this
reduction of freedom. In fact, the idea that ordinary consciousness is
not primary was previously  put forth by Alan Watts 
to describe what later became widely known as {\it mindfulness} \cite{livro_do_watts}
Indeed, it is possible to interpret the effects of Ayahuasca, and other psychedelics,  as being due to the temporary removal of the
  %\color{blue}
some of the restrictions that are necessary for sustaining ordinary (adult trained)
consciousness.
Without these restrictions, the mind reverts to the more flexible
state, in which self-referential narratives and thoughts about the
past or the future are no longer experienced as identical to the
reality that they are assumed to  represent \cite{livro_do_watts}.
%

% Mindfulness is also usually associated to meditative states \cite{Moore2009176}.
% Indeed, studies indicate some similar behaviors in functional
% complex network influenced by Ayahuasca and in meditation \cite{fernanda2014, brewer2011}. 
% In the primary state, it is easy to directly perceive, through the immediate
% experience, that ``the map is not the territory.''  
%
% Seen in this context, it is not entirely
% surprising that psychedelics and meditative practices may have therapeutic potential in addiction
% and depression \cite{Ramel12004,Carhart-Harris2016,Sanches2016,Bogenschutz01032015}.%,labate2013therapeutic}.

Relatively few studies have investigated entropy in brain functional
networks, hence
% Psylocibin
additional comments are in order. 
Tagliazucchi {\it et al.} \cite{tagliazicchi2014} showed that
psilocybin (psychedelic present in some species of mushrooms) may be
responsible for increases of a different entropy measure in functional
connectivity of the 4 regions of Default Mode Network (DMN), a
relevant functional network related to resting state.
%
% They shared the bold time series in several nonoverlap windows and valued partial
% correlation between 4 brain regions belonged to Default Mode Network (DMN, a relevant
% functional network related to resting state).
% They detect an increase of entropy, measured on distribution of repertory of connectivity
% motifs among these regions, to networks related to psilocybin influence when compered to networks
% from ordinary state and placebo.They do not value whole brain (probably due the computational
% cost restrictions of the method), but these results is a hint of increase of flexibility of panthers due the 
%increase of entropy on degree distribution of network of correlations
%
% Entropy and age
Recently, Yao {\it et al.} \cite{Yao2013a} correlated entropy increases in
the human brain with age. This study also supports the view that
entropy is correlated to {brain function (and perhaps also its
  development).}
%
% A global analysis of Shannon entropy for whole brain was performed by Yao et. al. \cite{Yao2013a}.
% They show entropy increases in functional weighted networks (matrix correlation) with the age in a sample
% of individuals from 6 to 76 years.  Here we can see that the entropy is in some way a function of brain development. 
%Propofol
Moreover,
in agreement with our results, 
% Along with our research,
a study by Schroter {\it et al.} \cite{schoter2012} similarly suggests that functional network topology may have a 
central role in consciousness quality. 
%
% A study by Schroter {\it et al.} \cite{schoter2012} may reinforce our results.
They investigated the effects on the human brain of the anesthetic
propofol, which can induce loss of consciousness \cite{Sarasso2015}.
They reported a decrease of the clustering coefficient, which is
strongly influenced by degree distribution (however, geodesic
distance remained unchanged).

We briefly comment on the limitations of our method:
 (i) the reduced number of subjects and the fact that all of them were 
experienced with Ayahuasca do not allow population inferences and do not 
elucidate whether the effects observed here were only due the acute 
administration or if previous experience also played a significant role; 
(ii) expectancy and suggestion were not controlled, as placebo was not used;
(iii) networks were built upon a number of critical choices, such as
the atlas used to partition the brain, the method used to build the
correlation matrix, and the cutoff thresholds for generating the
adjacency matrices from correlation matrices
\cite{Smith20121257,langer2013}, which may affect the final results;
 (iv) the chosen range of correlation values automatically limits the
networks'  behavior to a small-world network.
Despite this
limitation, it is important to highlight that several studies have
consistently demonstrated that brain networks
% in fact, 
exhibit a small-world behavior \cite{bassett2006}.

% Na verdade nós garantimos que as redes não sejam desconectadas e esparsas. O 
% critério utilizado é o mesmo sugerido para detectar redes small-world em redes reais. 
% Mas entendo que afirmarmos que estamos com esse critério garantindo que as
% redes sejam Small-World seja um pouco forte pq envolve uma discussão grande
% sobre definição de SW em redes reais. Eu sugiro a gente ficar fora dessa discussão. 
% Podemos em vez disso dizer nesse item que que a limitação está em selecionar somente redes conectadas e esparsas
% com um limiar comum a todos os sujeitos. Essa banda de correlação comum 
% pode fazer com que percamos informações, etc. Explicitamos a mesma limitação sem 
% assumir conjecturas.}

%
%
% Propofol is an anesthetic substance able to induce loss of consciousness. 
% Schroter et al. cite{schoter2012} evaluated global topological changes on functional 
% brain networks of humans under propofol effect. Unlike our results, they report decrease 
% of clustering coefficient in samples of subjects under propofol influence. Nevertheless, 
% there was not significative changes on geodesic distance. 	
% So, it suggest that propofol reduce local integration but maintain global integration. 

%

Finally, we speculate about  whether or not our finding of larger mean
geodesic distances may have any relation to self-reports of
``mind-expansion'' by users of psychedelics.
Could there be a direct relation between entropy increases and the
higher creativity reported by users of psychedelics? 
  Such questions merit further investigation.	
  In conclusion, our results are broadly consistent with the
  hypothesis that psychedelics increase the entropy in brain
  functions. By calculating the Shannon entropy of the degree
  distribution of complex networks generated from fMRI data, we have
  taken a new low-computational-cost approach to investigating brain
  function under the influence of psychedelics.

%

%\clearpage

\section*{Methods}\label{sec-methods}
%\bigskip\noindent{\bf Methods}\label{sec-methods}

\subsection*{Data acquisition and preprocessing}  \label{sec-data}
% \subsection{fMRI Acquisition}
The fMRI images were obtained in a 1.5 T scanner (Siemens, Magneton Vision), 
using an EPI-BOLD like sequence comprising 150 volumes,
with the following parameters: TR=1700 ms; TE=66 ms; FOV=220 mm; matrix 
64$\times$64; voxel dimensions of 1.72mm$\times$1.72mm$ \times $1.72 mm. 
It also was acquired whole brain high resolution T1-weighted images (156 
contiguous sagittal slices) using a multiplanar
reconstructed gradient-echo sequence, with the following parameters: TR=9.7 ms; 
TE=44 ms; flip angle 12$^{\circ}$; matrix 256$ \times $256; FOV= 256 mm,
voxel size$ = 1mm \times 1mm \times 1 mm$. %como esse último é usado?
The images were obtained from 10 healthy right-handed adult volunteers
(mean age 31.3, from 24 to 47 years), all who were experienced users of Ayahuasca with at
least 5 years use (twice a month) and at least 8 years of formal education.
The experimental procedure
was approved by the Ethics and Research
Committee of the University of São Paulo at Ribeirão Preto (process
number 14672/2006).
Written informed consent was obtained from all volunteers, who belonged to the Santo Daime religious organization.

Volunteers were not under medication for at least 3 months prior to
the scanning session and were abstinent from caffeine, nicotine and
alcohol prior to the acquisition.
They had no history of %encephalo cranial traumatism neither 
neurological or psychiatric disorders, as
assessed by DSM-IV structured interview \cite{american2000diagnostic}.
Subjects ingested 120-200 mL (2.2 mL/kg of body weight) of Ayahuasca
known to contain 0.8 mg/mL of DMT and $0.21$ mg/mL of harmine.
Harmaline was not detected via the chromatography analysis,
at the threshold of 0.02 mg/mL \cite{draulio2012}.
preprocessing steps were conducted in FSL (http://www.ndcn.ox.ac.uk/divisions/fmrib) and include: slice-timing correction, head motion correction and spatial smoothing (Gaussian kernel, FWHM = 5 mm). One volunteer was excluded from analysis due to excessive head movement (more than 3mm in some direction), leaving 9 participants (5 women) to our analysis. All images were spatially normalized to the Montreal Neurologic Institute (MNI152) \cite{Brett2002a} standard space, using a linear transformation. We also evaluated 9 regressors of non-interest using a General Linear Model (GLM): 6 regressors to movement correction, 1 to white matter signal, 1 to cerebrospinal fluid and 1 to global signal.
  Each volunteer was submitted to fMRI scanning under two distinct
  conditions: (i) before and (ii) 40 minutes subsequent to Ayahuasca
 intake.  In both cases, volunteers were in an awake resting state:
 they were requested to stay lying with eyes closed, without performing
 any task.
 One volunteer sample was excluded from analysis due to excessive head
 movement, leaving 9 participants (5 women) to our analysis.
% %

% Ten healthy volunteers were submitted to two distinct fMRI scanning
% sessions: (i) before and (ii) 40 minutes after Ayahuasca intake, when
% the subjective effects are noticeable.  In both cases, participants
% were instructed to close their eyes and remain awake and at rest,
% without performing any task.
%
% We thus have one dataset acquired from subjects under the influence of
% Ayahuasca, as well as a second control dataset for the same subjects
% in the ordinary state of wakeful rest.

%
% We performed a standard pre-processing on all samples of the fMRI
% data.  Details concerning data acquisition and pre-processing are
% described in appendix A.
%
% \subsection{Key concepts of complex networks}
%
%
%	

 \subsection*{Complex network metrics}
For a detailed overview of complex network theory, we refer readers to
refs. \cite{Bornholdt2003,Barabasi2002,rubinov2010}.
Each element of a network is known as a node (or vertex), and the
relation between a pair of nodes is represented by a connecting link
(or edge).  Links can have weights associated with them and can be
directed or undirected (or, equivalently bi-directional).
Nodes connected by a single link are known as nearest
neighbors~\cite{newman2010networks}.
Non-weighted undirected networks, i.e. those with symmetric and
unweighted links are isomorphic to a binary symmetric matrix known as the
adjacency matrix.
When a pair of nodes $i$ and $j$ are neighbors, the adjacency matrix
element is $a_{i,j}=1$ and $a_{i,j}=0$ otherwise.
Standard quantities of interest that help to characterize the topology
and complexity of networks \cite{onias2014,rubinov2010} include node
degree, geodesic distance, clustering coefficient, and local and
global network efficiencies.

Definitions:

(i) The degree $k_j$ of a node $j$ is the number of links that
it has with other nodes. The degree distribution of a network is
the normalized histogram of degrees over all nodes.

(ii) A geodesic path between two nodes is the shortest path from one
to the other, assuming such a path exists.  The geodesic distance
$d_{i,j}$ between nodes $i$ and $j$ is the number of links in the
geodesic path. If there is no such path, the geodesic distance is
defined as infinite. Given a network $G$ with $N$ nodes, the mean geodesic
distance is given by
% $
\begin{equation}
 D(G)=\frac{1}{N(N-1)}\sum_{i\neq j}{d_{i,j}} ~.
\label{distance}
% $  
 \end{equation}

(iii) The clustering coefficient quantifies the density of
triads of linked nodes, e.g., the fraction of the neighbors of a
node that are themselves neighbors. The clustering coefficient is defined
by
\begin{equation}
%  $ 
 C(G)=\frac{1}{N}\sum_{i \neq j \neq h } \frac{2}{k_i(k_i-1)}
    ~ a_{i,j}a_{j,h}a_{h,i} ~,
\label{cluster}
\end{equation}
% $
where $k_i$ is the degree of node $i$ and $a$ is the adjacency matrix element.

(iv) The efficiency, typically defined as the reciprocal of the
harmonic mean of geodesic distances, quantifies the influence of the
topology on flux of information through the network. Efficiency can be
global as well as  local.  We define global efficiency as
% % 
 \begin{equation}
% $
E_{\mbox{\tiny g}}(G)=\frac{1}{N(N-1)}\sum_{i \neq j \in G} \frac{1}{d_{i,j}}  ~,
 \end{equation}
% $
and local efficiency as
 \begin{equation}
% $
E_{\mbox{\tiny l}}
   (G)=\frac{1}{N}\sum_{i \in G} \bigg{(}\frac{1}{n_i(n_i-1)} \sum_{j \neq h \in g_i} \frac{1}{d_{h,j}} \bigg{)} ~,
 \end{equation}
% $
 %
 where $g_i$ are the subnetworks formed by neighbors of node $i$ and
 $n_i$ is the  number of nodes of this subnetwork \cite{latora2001}.
% %
% For calculate these measuraments we used Brain Connectivity Toolbox for Matlab .

In addition to these standard network properties, we also use the
Shannon entropy \cite{shannon1949} to quantify disorder or
uncertainty.
% % 
Specifically, we calculate the Shannon entropy functional of the
distribution of node degrees. Let $P$ be the normalized probability
distribution for node degree $k$, i.e.  $\sum_{k} P(k)=1$.  We define
the Shannon entropy $S[P]$ of the degree distribution $P(k)$ for a
network with $N$ nodes by:

 \begin{equation}
 S[P]= - \sum_{k} P(k)\log {P(k)} ~.
\end{equation}
Often the logarithm of base 2 is used \cite{cover2006elements} (e.g.,
in computer science), but we use the natural logarithm instead, so the
entropy values shown are in natural information units rather than in
bits.

\subsection*{Maslov algorithm for generating randomized networks}

Given $G$, one can select two non-overlapping pairs $(i,j)$ and
$(m,n)$ of linked nodes, then unlink them, and cross-link the
pairs according to $(i,m)$ and $(j,n)$. If this process is repeated
many times, the links become randomized, but the degree of each node
remains the same~\cite{Maslov2002}. Hence the entropy of the degree distribution is also
a conserved quantity.

\subsection*{Calculation of correlation matrix for brain regions}

% We use an anatomical approach to define network nodes.
We segmented the brain images into 110 brain regions according
to the Harvard-Oxford cortical and subcortical structural atlas
(threshold of $> 25\%$, using FMRIB Software Library, www.fmrib.ox.ac.uk/fsl).
Six regions had to be excluded from further analysis, as they were not sampled 
for all subjects, due to technical limitations during image acquisition.
For each of the 104 regions, an averaged fMRI time series was computed
from all voxels (a voxel is a 3D image block, analogous to the 2D
pixel). within that region using Marsbar (SPM toolbox). To reduce
confounders, we applied a band-pass filter ($\approx 0.03-0.07$ Hz)
using the maximum overlap wavelet transform (MODWT) with a Daubechies
wavelet to divide the signal into 4 scales of different frequency
bands. In keeping with the literature \cite{schoter2012, liuyong2008},
that point that resting state typically leads to low frequency
($\approx 0.01$ to $0.1$ Hz) \cite{Fransson2005}, we choose scale
3. We then calculated the Pearson correlation between these wavelet
coefficients from all possible pairs, thus obtaining a 104$\times$104
correlation matrix to represent each sample.
Only correlations with p $<$ 0.05 were considered.
%
% To reduce confounders we applied a band-pass filter ($\approx 0.03-0.07$ Hz) using the maximum overlap
% wavelet transform (MODWT) with a Daubechies wavelet to divide the signal into 4 scales of different frequency bands.
% In keeping with the literature \cite{schoter2012, liuyong2008}, that point that resting state 
% typically leads to low frequency ($\approx 0.01$ to $0.1$ Hz) \cite{Fransson2005},  we choose scale 3.
% %
% 
% For each of the 104 regions, a average of fMRI time series was computed, from signals of all voxels 
% \footnote{A voxel is a 3D image block, analogous to the 2D pixel.} within that region.
% %
% We calculated one time series from each region as the average of time series of all voxels
% \footnote{A voxel is a 3D image block, analogous to the 2D pixel. } for that region (using Marsbar, SPM toolbox).
% 
% Therefore, each fMRI image yields 104 time series, one for each
% cortical region.
% % % 
% % We then calculated the Pearson correlation between all possible pairs
% % of time series, thus obtaining a 104$\times$104 correlation matrix to
% % represent each sample. 
% % % 
%This correlation matrix was then used to define network links, whereas
%the nodes represented the cortical regions.
% 

%We aim map the topological structure of functional relations of
%cortical regions independent of nature of this relations.
%

\subsection*{Construction of complex networks from fMRI images}

A correlation matrix uniquely defines a weighted network.
Nonetheless, we are interested in generating non-weighted networks.
Hence, we need a function that maps correlation matrices to adjacency
matrices.
We use a thresholding function for this purpose.
Given a correlation matrix, we obtain the adjacency matrix by applying
a threshold to the absolute value of the elements of the correlation
matrix. Specifically, if the absolute value of the correlation matrix
element $|c_{i,j}|$ is larger than a defined threshold $\eta$, then a
link is assumed and the adjacency matrix element is taken to be 1
(i.e., $a_{i,j}=1$), while otherwise there is no link ($a_{i,j}=0$).
In order to obtain better statistics, we 
choose not a single value of $\eta$ but a range of values instead.
Then we analyze the behavior of the network properties over this
range. Using this approach, we create a number of networks for each
fMRI sample, all with the same number of nodes (104 nodes). For each
of these networks, we choose $\eta$ such that the density of links is
the same before and after Ayahuasca intake.
%
% The adjacency matrices, was constructed applying
% thresholds to absolute value of corrlation matrix: if the absolute value is larger
% than a defined threshold, a link is formed ($a_{i,j}=1$), otherwise no link is formed ($a_{i,j}=0$).
%
% Specifically, for all networks, we used the 104 cortical regions
% (yielding 104 nodes) and created the adjacent matrix to define links,
% by passing the correlation matrix through different thresholds and
% generating a binary adjacency matrix: for correlations larger than
% threshold the matrix element received 1 and 0 otherwise.
%

%  To better compare between the two conditions (before and after Ayahuasca
% intake), we created networks with the same number of nodes and density
% of connections.
%
%
We choose a range for the mean network degree to ensure the networks were
fully connected but also  sparse (to avoid random network behavior).
For this purpose, we adopt the following criteria: the network must
have lower global efficiency and greater local efficiency than its
randomized version. These criteria also ensure small-world behavior of
the networks \cite{Achard2007} (according to the definition of Watts
and Strogatz~\cite{WS1998}).
In order to obtain the same threshold range for all subjects, it is
necessary to exclude two of them from the analysis, since there is no
threshold range common between them and the other subjects.
Data from
a second subject was also excluded due to excessive head movement.
% (We hope to recover these 2 subjects in the future, but
% this will require the development of newer methods of data analysis.)
%
Following the criteria described above, the threshold range is  set to
$0.28 \leq \eta \leq 0.37$. We generate networks with mean degree in
the range $24 \leq \langle k\rangle \leq 39$. {We evaluate measures in
  degree increments of $\Delta \langle k \rangle=1$, thus obtaining 16
  different values of mean degree}.

In summary, we have 7 human subjects suitable for both conditions
(before and after ingestion).  The resulting sets of networks allow 16
different comparisons (i.e. of differing mean degrees) for each
subject before and after Ayahuasca ingestion.
% 
% Then, we calculate the topological and statistical properties of the
% all networks, which allowed us to compare the two conditions (i.e.,
% before and after Ayahuasca intake).
We calculate the topological measurements (using the Brain
Connectivity Toolbox for Matlab \cite{rubinov2010}).

\subsection*{Statistical testing}
Comparisons between the two conditions (i.e., before and after
Ayahuasca) are obtained from paired-sample Student's $t$-tests.
The $p$-values shown in some of the figures are as follows: values
$p<0.05$ in bold and $p<0.005$ indicated by asterisks (*).  The
implicitly assumed null hypothesis is that the difference of the
paired values are normally distributed with zero mean.

%

%

%

% \section*{Appendix A. Data acquisition and pre-processing steps}

%
%\subse

% Before generating the networks we performed a standard pre-processing
% of the fMRI data \cite{huettel2004functional}, including slice-timing
% correction, head motion correction and spatial smoothing (Gaussian
% kernel, FWHM = 5 mm). We evaluated 9 regressors using a General Linear
% Model (GLM): 6 regressors to movement correction, 1 to white matter
% signal, 1 to cerebrospinal fluid and 1 to global signal.
%
% The processing was made using FSL Software \cite{Woolrich2009S173,Smith2004S208}, 
% a free library of statistical tools available by Oxford Centre
% for Functional MRI of the Brain ({http://www.ndcn.ox.ac.uk/divisions/fmrib}).
%
% All images were spatially normalized according a anatomical standard space, the Montreal
% Neurologic Institute (MNI152 template) \cite{Brett2002a} using a linear
% transformation. 
% %
% 

\nolinenumbers\relax % Commence numbering lines

\bibliographystyle{nature}

\clearpage

%\section*{References}

\bibliography{entropy_aya10}

\clearpage

  \section*{Acknowledgements}

We thank Santo Daime members for volunteering and for providing the
Ayahuasca.  We thank Sidarta Ribeiro for discussions,
José~C.~Cressoni, Marco~A.~A.~da~Silva, and Carlos~Viol for feedback
and CAPES and CNPq for funding.
%
% (colocar mais alguém da banca?).  
%We thank
AV thanks UFV and Science without
Borders (CAPES Grant No. 88881.030375/2013-01) for funding 
and Guillermo Cecchi and Irina Rish for their  hospitality 
and discussions during her year at IBM.

  \section*{Author contributions}

D.B.A. recruited the volunteers for data acquisition and conceived the
study.  A.V., F.P.-F. and H.O performed fMRI data preprocessing, complex
network construction and evaluated standard network features.
A.V. and G.M.V. performed complex network analysis and statistical
analysis.  All authors contributed equally to the final overall design
of the study.

  \section*{Competing financial interests}

The authors declare no competing financial interests.

%\end{document}


  \clearpage
\begin{center}
\centerline{(a)  \mbox{~~~~~~~~~~~~~~~~~~~~~~~~~~~~~~~~~~~~~~~~~~~~~~~~~~~~~~~~~~~~~~~}  (b)}  
  \centerline{
\includegraphics[scale=0.34]{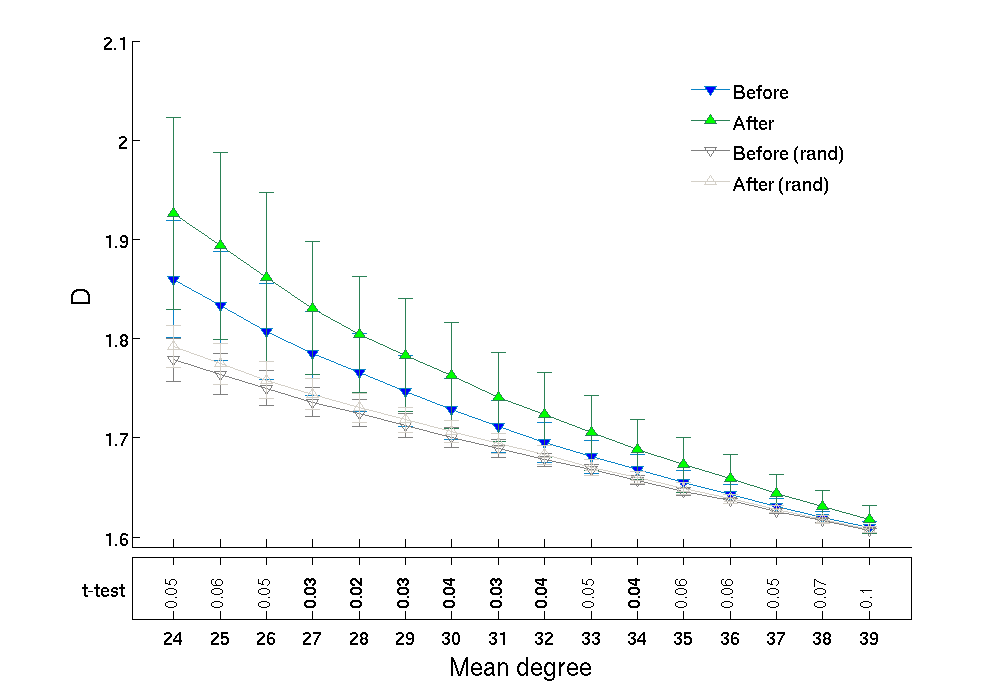}
\includegraphics[scale=0.34]{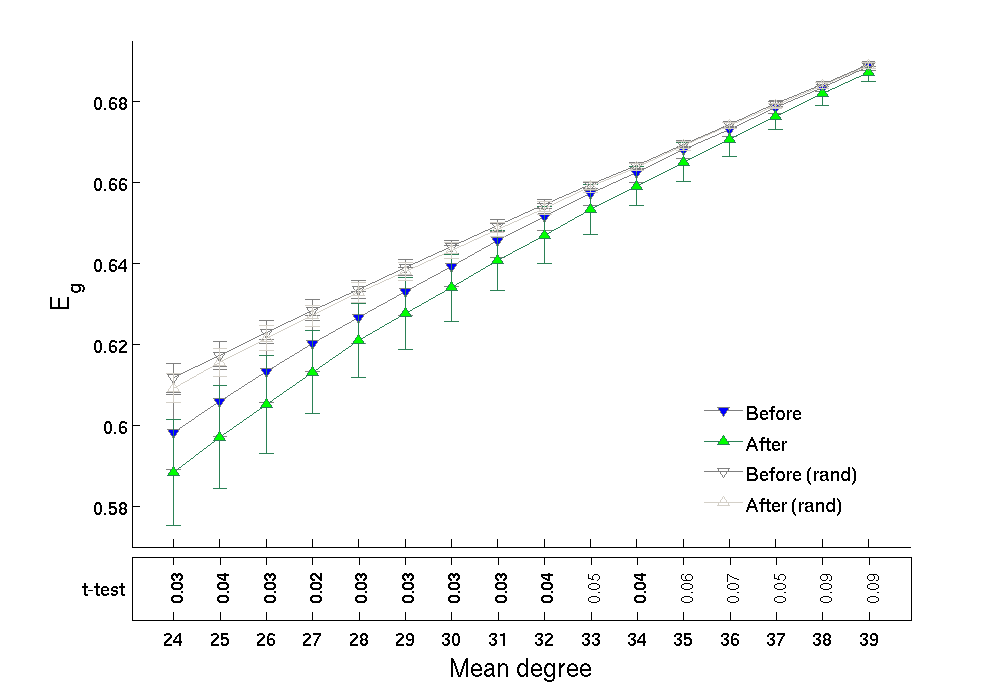}
  }

\medskip

\centerline{(c)  \mbox{~~~~~~~~~~~~~~~~~~~~~~~~~~~~~~~~~~~~~~~~~~~~~~~~~~~~~~~~~~~~~~~}  (d)}
  \centerline{  
\includegraphics[scale=0.34]{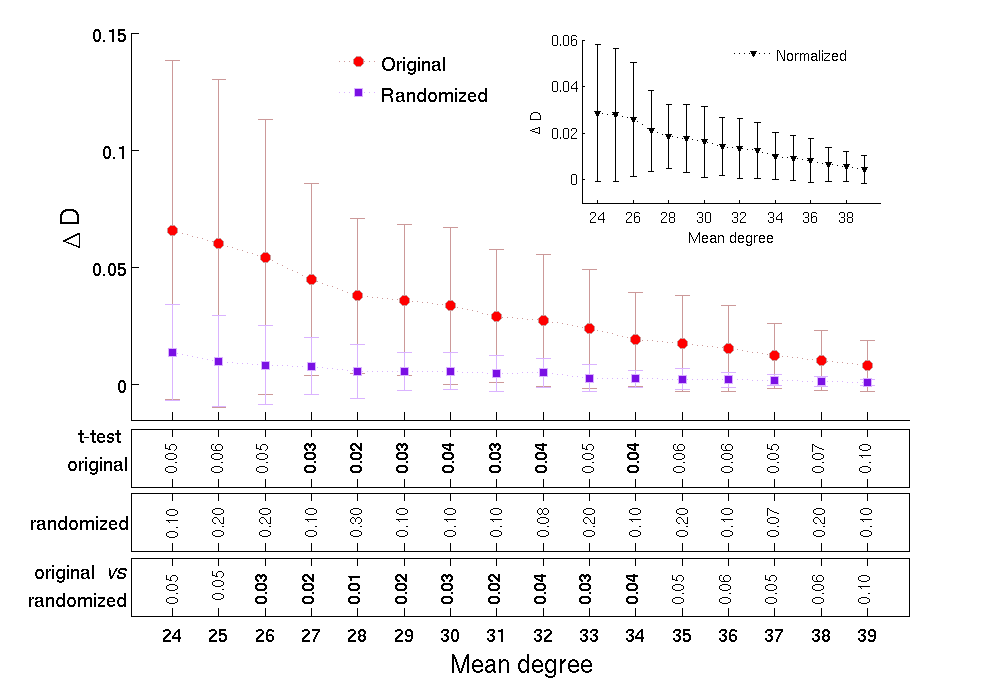}
\includegraphics[scale=0.34]{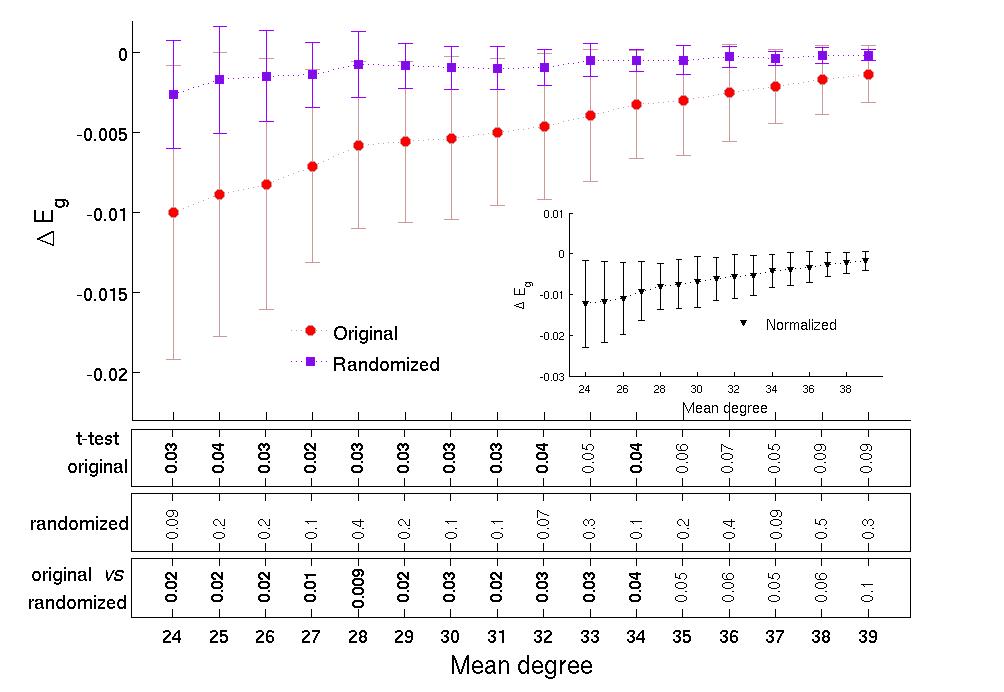}  
  }

  \medskip

\centerline{(e)  \mbox{~~~~~~~~~~~~~~~~~~~~~~~~~~~~~~~~~~~~~~~~~~~~~~~~~~~~~~~~~~~~~~~}  (f)}    
  \centerline{
\includegraphics[scale=0.33 ]{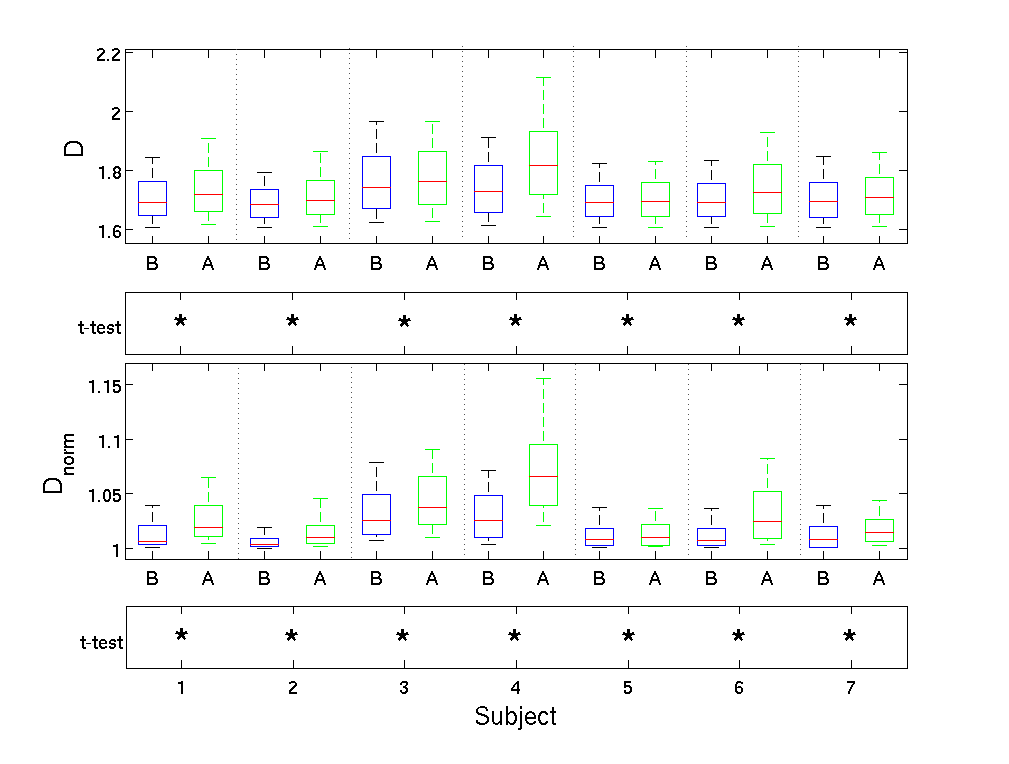}
\includegraphics[scale=0.33]{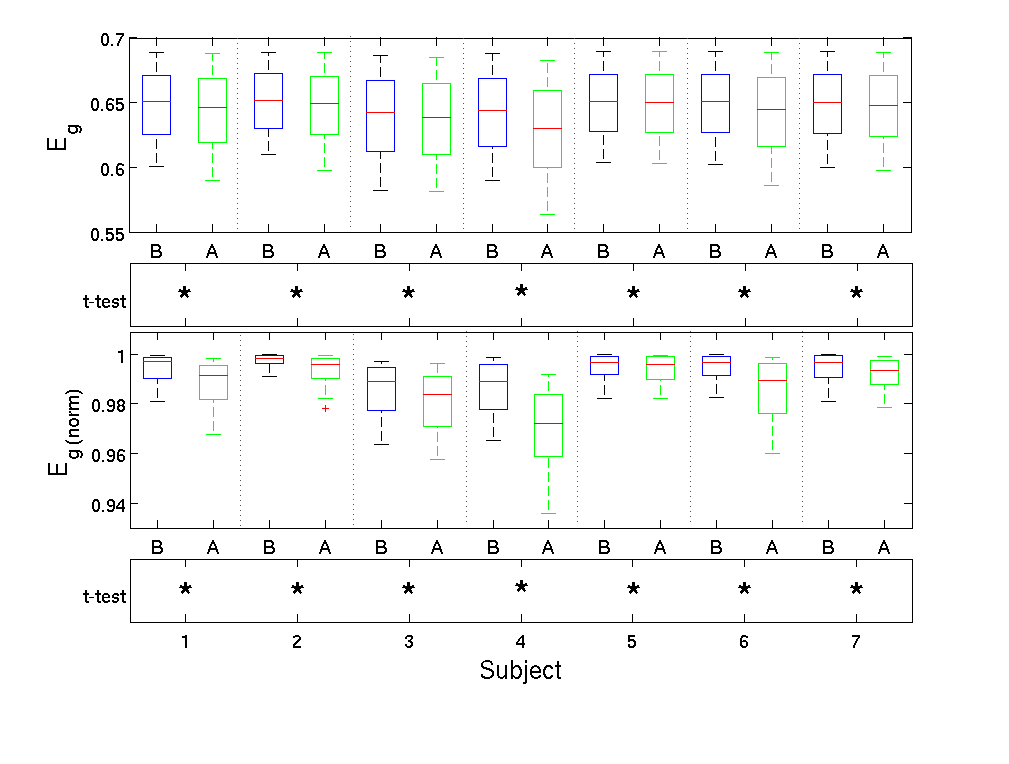}  
}

\vspace{-4mm}
  \noindent Figure 5.  (See next page for caption.)

\end{center}

  \clearpage

  \begin{figure}[h]
%  \begin{center}
    
  \caption{{\bf Global efficiency and integration decrease.} Geodesic
    distance $D$ (left column) and global efficiency $E_{\mbox{\tiny
        g}}$ (right column). Plots (a) and (b) show means $\pm$ 1
    standard deviations, calculated from the complex networks of all 7
    subjects, as well as from their corresponding iso-entropic
    randomized networks, for 16 different mean degrees.
    Plots (c) and (d) show the change in $D$ and $E_{\mbox{\tiny
        g}}$ after Ayahuasca ingestion.
    The inset shows normalized values (see text).
    Boxplots (e) and (f) show the same information, subject-by-subject.
    As in previous figures, the rows below the plots show $p$-values for the $t$-test,
    with asterisks (*) indicating $p<0.005$.
  }
  \label{distance}
%  \end{center}
  \end{figure}  

  \clearpage

  \begin{center}
\centerline{(a)  \mbox{~~~~~~~~~~~~~~~~~~~~~~~~~~~~~~~~~~~~~~~~~~~~~~~~~~~~~~~~~~~~~~~}  (b)}    
    \centerline{
\includegraphics[scale=0.35]{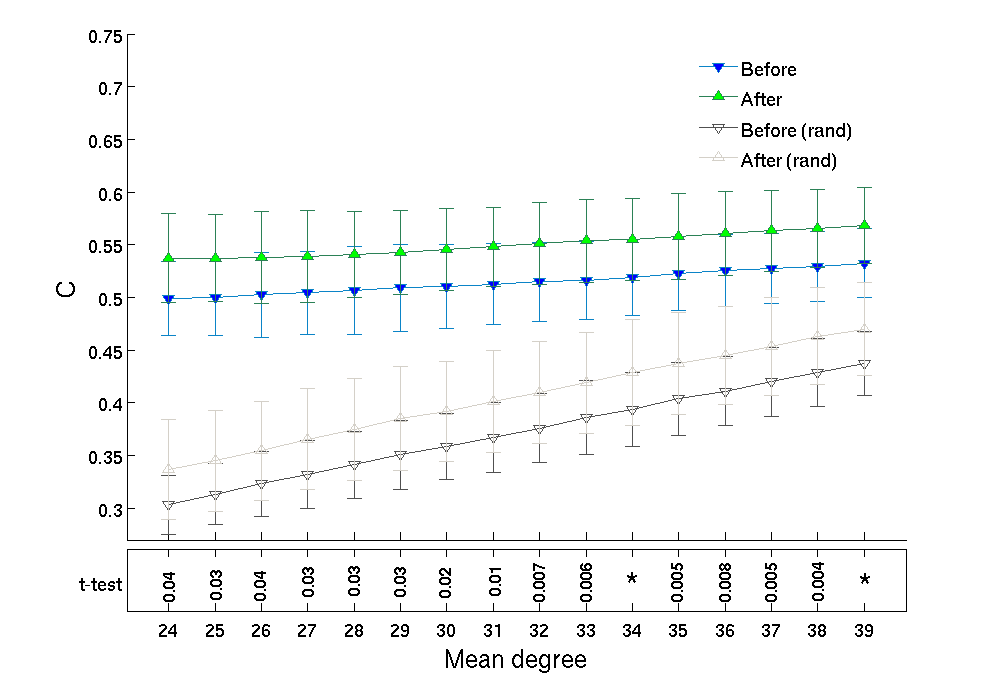}
\includegraphics[scale=0.35]{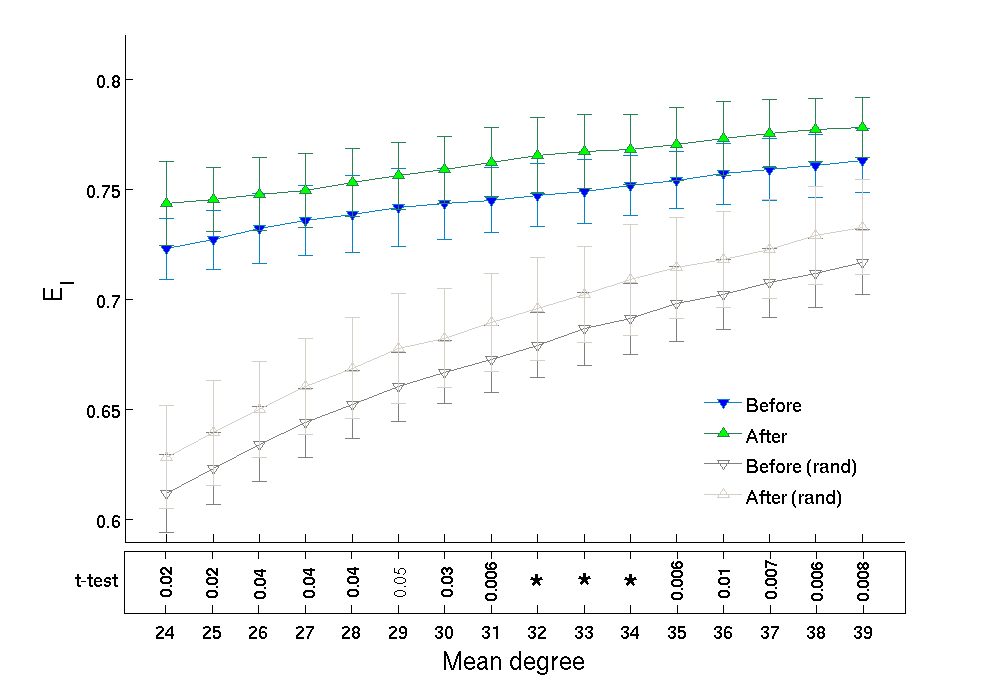} 
}
\medskip

\centerline{(c)  \mbox{~~~~~~~~~~~~~~~~~~~~~~~~~~~~~~~~~~~~~~~~~~~~~~~~~~~~~~~~~~~~~~~}  (d)}    
    \centerline{
      \includegraphics[scale=0.33]{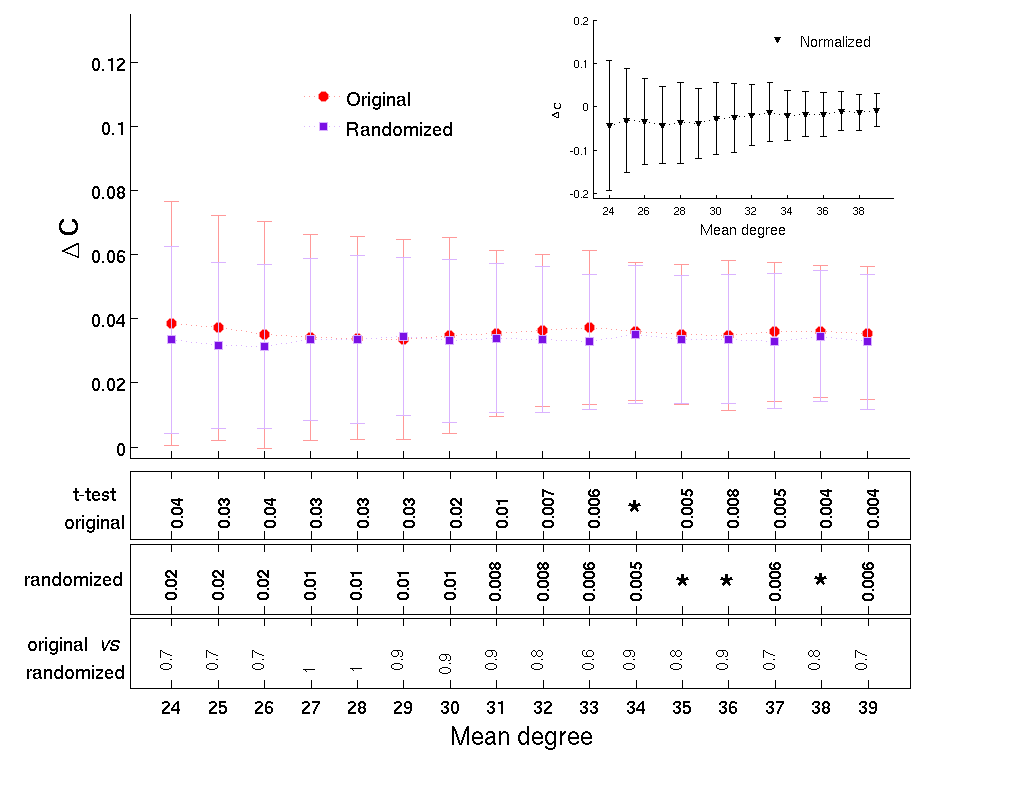}
     \raisebox{5mm}{ \includegraphics[scale=0.35]{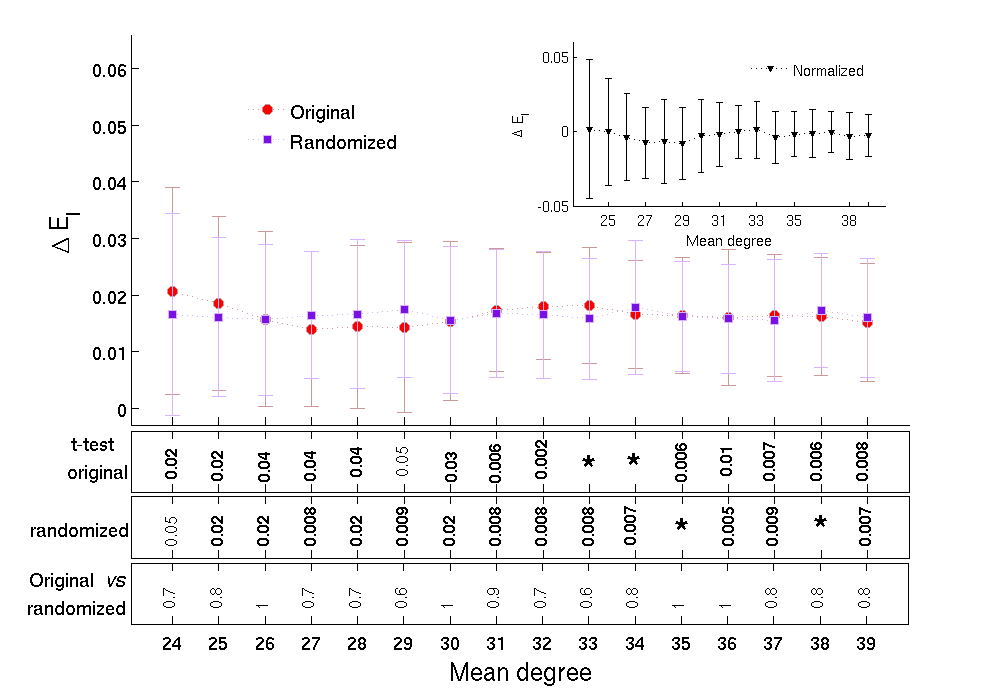}}
}
    \smallskip

%   \vspace{-2mm}
    
\centerline{(e)  \mbox{~~~~~~~~~~~~~~~~~~~~~~~~~~~~~~~~~~~~~~~~~~~~~~~~~~~~~~~~~~~~~~~}  (f)}        
    \centerline{
\includegraphics[scale=0.34]{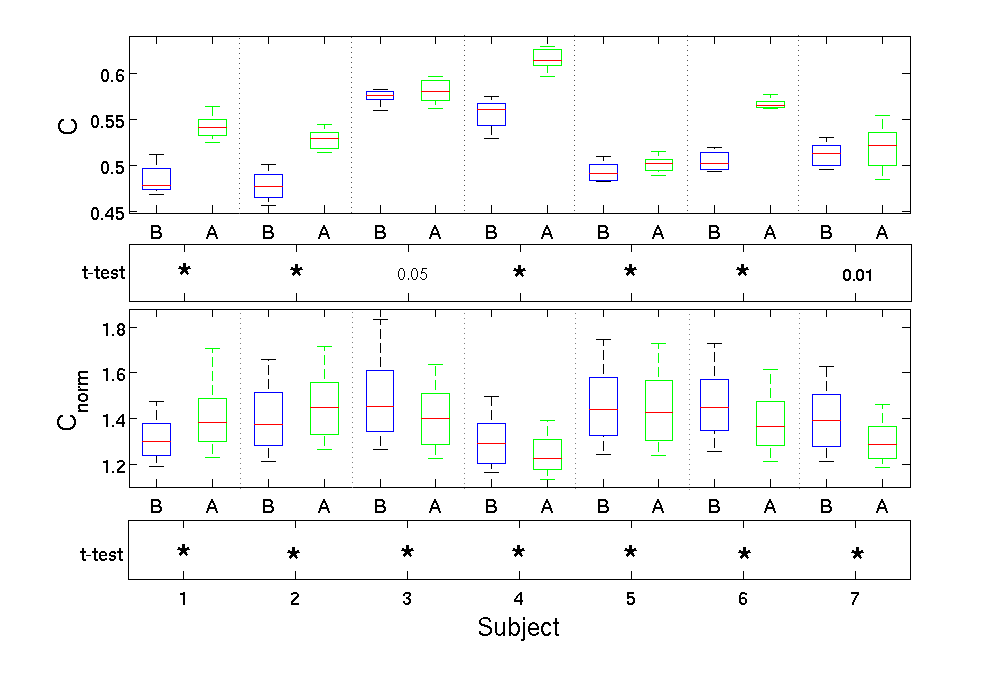}
\includegraphics[scale=0.34]{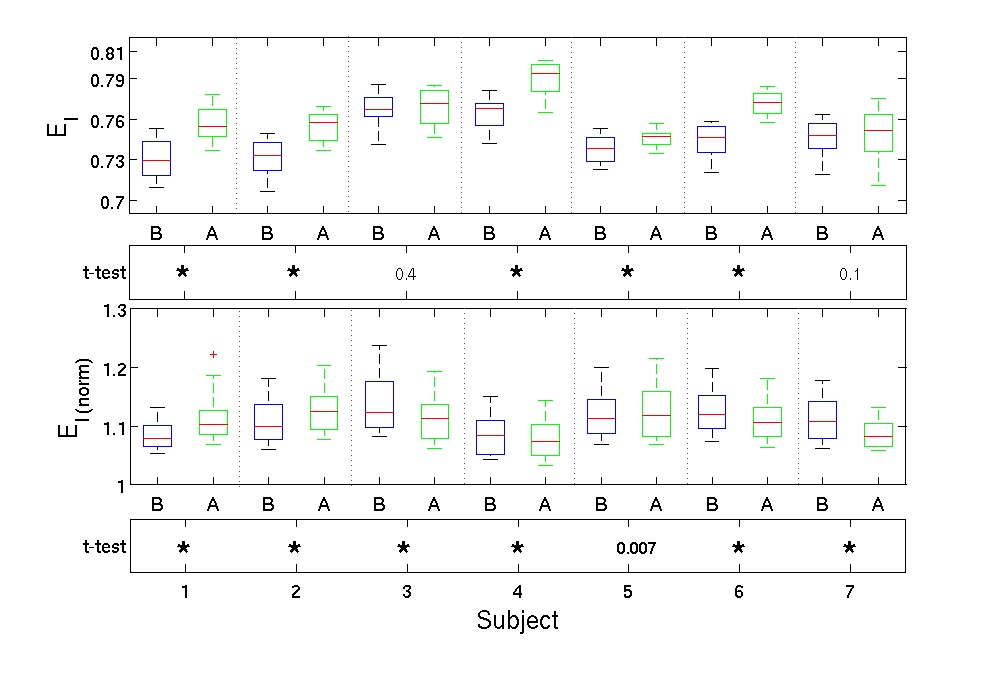}
    }
    \vspace{-7mm}\noindent Figure 6.  (See next page for caption.)    
  \end{center}

\clearpage

  \begin{figure}[ht]
  \caption{ {\bf Local efficiency and integration increase.} Clustering coefficient $C$ (right
    column) and local efficiency $E_{\mbox{\tiny l}}$ (left column).
    Plots (a) and (b)  show means $\pm$ 1 standard
    deviations, calculated from the complex networks of all 7
    subjects, as well as from their corresponding iso-entropic
    randomized networks, for 16
    different mean degrees.
    Plots (c) and (d) show the change in $C$ and $E_{\mbox{\tiny
        l}}$ after Ayahuasca ingestion.
    The inset shows normalized values (see text).
    Boxplots (e) and (f)  show the same information, subject-by-subject.
    As before, the rows below the plots show $p$-values for the $t$-test,
    with asterisks (*) indicating $p<0.005$.
  }
  \label{clustering}
  \end{figure}  

\end{document}